\documentclass[reprint]{aastex701}
\usepackage{amsmath,amssymb,mathtools,bm}
\usepackage{graphicx}
\usepackage{epstopdf}
\usepackage{xcolor}
\usepackage{array}
\usepackage{booktabs,tabularx,microtype}
\newcommand{\dd}{\mathrm{d}}
\newcommand{\e}{\mathrm{e}}
\newcommand{\fRT}{f(R,T)}
\newcommand{\Schw}{\mathrm{Schw}}
\newcommand{\PINN}{\mathrm{PINN}}
\newcommand{\Cheb}{\mathrm{Cheb}}
\newcommand{\tot}{\mathrm{tot}}
\newcommand{\eq}{\mathrm{eq}}
\newcommand{\coup}{\mathrm{coup}}
\newcommand{\sm}{\mathrm{sm}}
\newcommand{\minH}{\mathcal{H}_{\min}}
\newcommand{\Ag}{\mathcal{A}_g}
\newcommand{\Ohat}{\hat{\Omega}}
\setcitestyle{numbers,square}
\makeatletter
\def\frontmatter@title@above{}
\journalinfo{}
\makeatother

\shorttitle{Odd-parity QNMs in trace-quadratic \fRT{} gravity}

\begin{document}

\title{Odd-parity perturbations of trace-quadratic $f(R,T)$ black holes with anisotropic matter: admissible branches, axial ringdown, and a coupled-PINN benchmark}

\correspondingauthor{Mushtaq Ahmad}

\author{Mushtaq Ahmad}
\email{mushtaq.sial@nu.edu.pk}\affiliation{National University of Computer and
Emerging Sciences,\\ Chiniot-Faisalabad Campus, Pakistan.}
\author{M. Farasat Shamir}
\email{farasat.shamir@nu.edu.pk}

\affiliation{National University of Computer and Emerging Sciences, Lahore Campus, Pakistan.}
\author{Adnan Malik}
\email{adnan.malik@zjnu.edu.cn; adnan.malik@skt.umt.edu.pk; adnanmalik_chheena@yahoo.com}
\affiliation{School of Nuclear Science and Technology, University of South China, Hengyang 421001, China}
\affiliation{School of Mathematics and Physics, University of South China, Hengyang, 421001, China.}
\affiliation{Department of Mathematics, University of Management and Technology, Sialkot Campus, Pakistan.}
\author{Ahdab K. Althukair}
\email{akalthukair@pnu.edu.sa}
\affiliation{Department of Physics, College of Sciences, Princess Nourah bint Abdulrahman University, P.O. Box 84428, Riyadh 11671, Saudi Arabia.}
\begin{abstract}
We study odd-parity gravitational perturbations of static black holes in trace-quadratic $f(R,T)=R+\alpha T^2$ gravity supported by an anisotropic effective fluid with constant closure parameters $(w_r,w_t)$. From the unreduced axial system and its principal symbol, we identify the sector of parameter space that supports a regular horizon, asymptotic flatness, and hyperbolic odd-sector evolution. Within this closure the admissible branch lies at negative $w_r$, while the commonly used positive-$w_r$ family fails the background regularity test and is kept only as a numerical comparison branch. On static admissible backgrounds the odd sector is exactly equivalent to Einstein gravity coupled to a frozen effective anisotropic fluid, so the physical axial spectrum is governed by a single gauge-invariant master equation. For the anchored branch $(w_r,w_t)=(-0.2,0.15)$ we compute the fundamental axial $\ell=2$ quasinormal mode with an exact Chebyshev solve. The mass-normalized spectrum differs from Schwarzschild by about $22\%$, whereas no statistically resolved direct $\alpha$-dependence appears within the conservative spectral envelope over $0\le \alpha/M^2\le 0.3$. We also construct a coupled physics-informed neural network for the unreduced two-field eigenproblem and use it to benchmark the inadmissible comparison branch. A closure-level audit of the anchored family shows positive diagnostic combinations associated with the null, weak, and dominant energy conditions, denominator safety in the modified balance law, and an effective exterior mass fraction of about $20\%$, while indicating that the constant-$(w_r,w_t)$ model should be read as an effective anisotropic stress rather than as a microphysical fluid. Within this closure, the main observable imprint in axial ringdown comes from the existence of the matter-supported branch itself, not from direct variation of the trace coupling.
\end{abstract}


\section{Introduction}
\label{sec:intro}

The gravitational-wave detections by the LIGO--Virgo--KAGRA network have turned black-hole ringdown into an observational probe of strong-field gravity rather than a purely theoretical construct \citep{Aasi2015,Acernese2015,Abbott2016,Abbott2019,Abbott2021,Abbott2023}. In the linear regime, the signal is described by a superposition of damped quasinormal modes (QNMs), whose complex frequencies encode both the background spacetime and the dynamical content of the underlying theory \citep{Berti2018,Dreyer2004,Berti2007,Kokkotas1999,Berti2009,Konoplya2011}. Reliable QNM calculations are therefore essential for black-hole spectroscopy and for using ringdown data to test extensions of general relativity (GR).

Among modified-gravity models, theories with direct geometry--matter couplings are especially interesting because they can alter both the background solution and the perturbation equations. In \fRT{} gravity, introduced in the original trace-coupled proposal \citep{Ref_fRT}, the action depends on the Ricci scalar $R$ and on the trace $T$ of the stress tensor. The theory is therefore sensitive to the matter action through $\Theta_{\mu\nu}$ and obeys a modified matter-balance law. For black-hole spacetimes, this opens the possibility of matter-supported exterior structure and ringdown spectra that differ from the vacuum GR prediction. Analytic solutions and related phenomenology in \fRT{} settings have already been studied \citep{Santos2023,Tangphati2024,Mohan2025}, but a systematic frequency-domain treatment of the coupled odd-parity gravitational sector has remained unavailable.

The main technical point is that the unreduced Regge--Wheeler-gauge axial system of the quadratic model is genuinely coupled. In GR, and in many beyond-GR examples, odd-parity perturbations can be written directly as a single second-order wave equation with an effective potential, which makes shooting, continued-fraction, spectral, or WKB methods comparatively direct \citep{ReggeWheeler1957,Chandrasekhar1983Book,SchutzWill1985,IyerWill1987,Iyer1987,ChandrasekharDetweiler1975,Leaver1985,Nollert1993,Jansen2017,Chung2024,BlazquezSalcedo2024Kerr}. Here, after harmonic decomposition, two gravitational amplitudes remain coupled to an axial matter amplitude that can be removed only through the modified divergence law. That unreduced system is the natural arena for the PINN and for the principal-symbol admissibility analysis. On static branches, however, the same odd sector can also be rewritten exactly as Einstein gravity with a frozen effective anisotropic fluid, so the physical axial spectrum is governed by a single gauge-invariant master equation. The two descriptions play different roles in what follows: the unreduced system is used for benchmarking and admissibility diagnostics, while the exact master equation provides the production route on the admissible branch.

Physics-informed neural networks (PINNs) are well suited to this setting. They represent the unknown functions with neural networks and enforce the differential equations directly at collocation points \citep{Raissi2019,Nascimento2020}. PINN calculations of black-hole QNMs already exist for single-field problems in GR and in modified gravity, especially when the perturbation equations reduce to one-dimensional master equations \citep{Cornell2022,Cornell2024,Luna2023,Ovgun2021,Ncube2021,Patel2024}. The problem here is more demanding because the background is numerical and matter supported, and because the odd sector remains genuinely coupled. The central question is not whether a PINN can reproduce a known master-equation spectrum, but whether it can solve a coupled gravitational eigenproblem with exact horizon regularity, nontrivial asymptotics, and a physically meaningful admissibility boundary in parameter space.

Recent work helps locate the present contribution. PINN-based QNM extraction has now been demonstrated on numerical modified-gravity backgrounds in Einstein-scalar-Gauss--Bonnet theory \citep{Luna2024PINN}. High-accuracy spectral perturbation methods can compute gravitational QNM corrections without first decoupling the full set of linearized field equations, at least perturbatively around spinning modified-gravity black holes \citep{ChungYunes2024}. On the odd-parity side, Horndeski ringdown has recently been revisited with time-domain validation and phenomenological error estimates \citep{Yang2025Horndeski}, while coupled odd-parity two-field systems have been analyzed in vector-tensor effective field theory \citep{Tomizuka2025VectorTensor}. Axial gravitational QNMs have also been computed for curvature-quadratic black holes with extra spin-two degrees of freedom \citep{Antoniou2025Quadratic}. Against that backdrop, this paper addresses a distinct regime: a trace-quadratic geometry--matter coupling, a matter-supported exterior, and an explicit admissibility-first interpretation of the axial spectrum.

In this work we derive the odd-parity perturbation system for a matter-supported black hole in trace-quadratic $f(R,T)=R+\alpha T^2$ gravity and use it to identify which anisotropic-fluid closures support a regular, asymptotically flat exterior. We then focus on the anchored admissible branch and show that its axial $\ell=2$ fundamental mode is numerically stable, with no statistically resolved direct $\alpha$-dependence within the conservative spectral envelope over $0\le \alpha/M^2\le 0.3$. A coupled PINN solver is developed and tested on the associated two-field eigenproblem, while the admissible branch itself is computed from an exact solve of the gauge-invariant axial master equation on the self-consistent anchored backgrounds. Within the constant-$(w_r,w_t)$ closure studied here, axial ringdown is therefore controlled far more strongly by the existence of the matter-supported branch than by direct variation of $\alpha$ along that branch.

The rest of the paper is organized as follows. Section~\ref{sec:model} develops the trace-quadratic \fRT{} model, the anisotropic-fluid prescription, the odd-parity perturbation equations, and the background admissibility constraints. Section~\ref{sec:numerics} summarizes the numerical strategy, including the coupled PINN, the Chebyshev methods, the hyperbolicity screen, and the uncertainty model. Section~\ref{sec:results} presents the admissible branch and the positive-$w_r$ comparison scan. Section~\ref{sec:discussion} discusses interpretation, limitations, and the polar-sector outlook. Technical derivations and auxiliary comparison catalogues are collected in the appendices.

\section{Trace-quadratic \fRT{} gravity and the axial perturbation problem}
\label{sec:model}

\subsection{Field equations and matter prescription}
\label{subsec:fieldeqs}

We start from the action
\begin{equation}
S=\frac{1}{16\pi}\int \dd^4x\,\sqrt{-g}\,f(R,T)+\int \dd^4x\,\sqrt{-g}\,L_m^{\mathrm{eff}},
\end{equation}
with quadratic trace coupling
\begin{equation}
f(R,T)=R+\alpha T^2.
\end{equation}
Because \fRT{} gravity depends explicitly on the matter action through $\Theta_{\mu\nu}$, the matter sector must be specified explicitly. We adopt the effective anisotropic-fluid Lagrangian
\begin{equation}
L_m^{\mathrm{eff}}=p_t-\frac{1}{2}(\rho+p_t)(g^{\alpha\beta}u_\alpha u_\beta+1)-\frac{1}{2}(p_r-p_t)(g^{\alpha\beta}\chi_\alpha\chi_\beta-1),
\end{equation}
where $u_\mu u^\mu=-1$, $\chi_\mu\chi^\mu=1$, and $u_\mu\chi^\mu=0$. The corresponding stress tensor is
\begin{equation}
T_{\mu\nu}=(\rho+p_t)u_\mu u_\nu+p_tg_{\mu\nu}+(p_r-p_t)\chi_\mu\chi_\nu.
\end{equation}
For this matter prescription,
\begin{equation}
\Theta_{\mu\nu}=-2T_{\mu\nu}+p_t g_{\mu\nu}.
\end{equation}
Since $f_R=1$ identically, all derivative terms involving $f_R$ vanish and the field equations reduce to
\begin{equation}
G_{\mu\nu}=(8\pi-2\alpha T)T_{\mu\nu}+\alpha\left(2p_tT+\frac{1}{2}T^2\right)g_{\mu\nu}.
\label{eq:einsteinlike}
\end{equation}
Taking the divergence and using $\nabla^\mu G_{\mu\nu}=0$ yields the modified matter-balance law
\begin{equation}
\nabla_\mu T^{\mu}{}_{\nu}=\frac{\alpha}{4\pi-\alpha T}\left[T^{\mu}{}_{\nu}\,\partial_\mu T-T\,\partial_\nu p_t-\left(p_t+\frac{1}{2}T\right)\partial_\nu T\right].
\label{eq:modbalance}
\end{equation}
All results below refer to this specific matter prescription; different matter actions would realize the same formal \fRT{} Lagrangian differently.

\subsection{Background equations, solver workflow, and consistency constraints}
\label{subsec:background}

For static spherical symmetry we write the metric as
\begin{equation}
\dd s^2=-N(r)\sigma(r)^2\dd t^2+\frac{\dd r^2}{N(r)}+r^2\dd\Omega_2^2,
\label{eq:Nsigma_metric}
\end{equation}
with
\begin{equation}
N(r)=e^{-2\nu_0(r)},
\qquad
\sigma(r)=e^{\mu_0(r)+\nu_0(r)},
\qquad
e^{2\mu_0}=N\sigma^2.
\end{equation}
The background matter is specified by the constant phenomenological closure
\begin{equation}
p_{r0}=w_r\rho_0,
\qquad
p_{t0}=w_t\rho_0,
\qquad
T_0=q\rho_0,
\qquad
q\equiv -1+w_r+2w_t.
\end{equation}
Defining
\begin{equation}
\mathcal{C}_N\equiv \frac{1}{2}\left(w_r^2+8w_rw_t+2w_r+12w_t^2-3\right),
\end{equation}
Eqs.~\eqref{eq:einsteinlike} and \eqref{eq:modbalance} become the first-order system
\begin{align}
N'&=\frac{1-N}{r}-8\pi r\rho_0+\alpha\mathcal{C}_N\,r\rho_0^2,
\label{eq:Nprime}
\\
\frac{\sigma'}{\sigma}&=\frac{r(1+w_r)\rho_0\left(4\pi-\alpha q\rho_0\right)}{N},
\label{eq:sigmaprime}
\\
\rho_0'&=-\frac{\rho_0\left[(1+w_r)\left(\frac{N'}{2N}+\frac{\sigma'}{\sigma}\right)+\frac{2(w_r-w_t)}{r}\right]}{w_r-\dfrac{\alpha q\rho_0(1+w_r-6w_t)}{2\left(4\pi-\alpha q\rho_0\right)}}.
\label{eq:rhoprime}
\end{align}
The angular Einstein equation is then a consistency check rather than an independent evolution equation.

Equations~\eqref{eq:Nprime}--\eqref{eq:rhoprime} fully specify the background solver. A practical implementation is:
\begin{enumerate}
\item Fix the horizon scale by setting $r_H=1$ during the integration. The physical mass is recovered afterwards from the asymptotic metric coefficient.
\item Start at $r=r_H(1+\varepsilon)$ with $\varepsilon\ll1$ using the near-horizon expansion given below. A constant rescaling of time lets one set $\sigma_H=1$ initially and renormalize to $\sigma(\infty)=1$ after the integration.
\item Integrate the ODE system outward with an adaptive solver. A shooting parameter such as the leading matter amplitude $\rho_s$ is adjusted until the solution joins the decaying branch and the asymptotic metric is flat within numerical tolerance.
\item Extract $M$ from the asymptotic fit of $N(r)$ and verify a posteriori that the angular Einstein equation and the modified balance law are satisfied along the numerical profile.
\end{enumerate}

The same ODE system also yields the consistency conditions. Let $r=r_H+\Delta r$ near the horizon and assume a nontrivial matter-supported branch of the form
\begin{equation}
N(r)=\frac{\Delta r}{r_H}+\cdots,
\qquad
\sigma(r)=\sigma_H+\cdots,
\qquad
\rho_0(r)=\rho_s\,\Delta r^{s}+\cdots.
\end{equation}
Equation~\eqref{eq:rhoprime} then gives the indicial relation
\begin{equation}
s=-\frac{1+w_r}{2w_r}.
\label{eq:s_index}
\end{equation}
A nontrivial decaying matter branch therefore requires $w_r<0$; otherwise $s\le0$ and the matter either diverges or fails to vanish at the horizon. Requiring $\sigma'/\sigma$ to remain finite gives the stronger condition $s\ge1$, hence
\begin{equation}
-\frac{1}{3}\le w_r<0.
\label{eq:wr_constraint}
\end{equation}
The illustrative positive-$w_r$ case $w_r=+0.2$ therefore does not belong to the regular matter-supported sector of the stated model.

At large radius, let
\begin{equation}
\rho_0(r)\sim \rho_\infty r^{-n_\infty},
\qquad
T_0(r)\sim q\rho_\infty r^{-n_\infty},
\qquad r\to\infty.
\end{equation}
The leading terms of Eq.~\eqref{eq:rhoprime} imply
\begin{equation}
n_\infty=\frac{2(w_r-w_t)}{w_r}.
\label{eq:ninfty}
\end{equation}
Finite ADM mass requires $n_\infty>3$, so in the admissible $w_r<0$ sector one must also have
\begin{equation}
w_t>-\frac{w_r}{2}.
\label{eq:wt_constraint}
\end{equation}
The asymptotic metric then takes the form
\begin{equation}
N(r)=1-\frac{2M}{r}+O\!\left(r^{2-n_\infty}\right),
\qquad
\sigma(r)=1+O\!\left(r^{2-n_\infty}\right),
\label{eq:asymptotics_corrected}
\end{equation}
with $n_\infty>3$. A falloff of the form $T_0\sim Q_T/r^3$ is therefore incompatible with asymptotic flatness in this closure.

\begin{figure}[t]
\centering
\includegraphics[width=0.92\columnwidth]{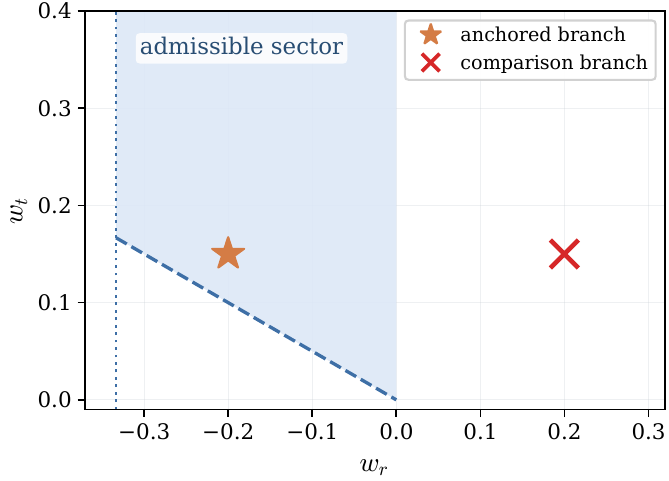}
\caption{Analytic admissibility wedge for the constant-$(w_r,w_t)$ closure. The shaded region satisfies the horizon-regularity condition $-1/3\le w_r<0$ and the finite-mass condition $w_t>-w_r/2$. The star marks the anchored admissible branch used for physical inference, while the cross marks the positive-$w_r$ comparison family retained only for numerical benchmarking.}
\label{fig:admissibility_wedge}
\end{figure}

\begin{table}[t]
\centering
\small
\setlength{\tabcolsep}{4pt}
\renewcommand{\arraystretch}{1.08}
\caption{Closure-level audit for the anchored admissible branch $(w_r,w_t)=(-0.2,0.15)$ with $\rho_0>0$ and $\alpha\ge0$. The listed combinations are diagnostic statements within the effective closure. The final row records the formal barotropic surrogates $c_{s,r}^2=\partial p_r/\partial\rho$ and $c_{s,t}^2=\partial p_t/\partial\rho$; because the closure is effective, they are not interpreted as literal microphysical propagation speeds.}
\label{tab:closure_audit}
\begin{tabular}{ll}
\toprule
\textbf{Diagnostic} & \textbf{Anchored branch value} \\
\midrule
radial NEC/WEC & $\rho+p_r=0.8\,\rho_0>0$ \\
tangential NEC/WEC & $\rho+p_t=1.15\,\rho_0>0$ \\
radial DEC & $\rho-|p_r|=0.8\,\rho_0>0$ \\
tangential DEC & $\rho-|p_t|=0.85\,\rho_0>0$ \\
SEC & $\rho+p_r+2p_t=1.10\,\rho_0>0$ \\
balance denominator & $4\pi-\alpha q\rho_0=4\pi+0.9\,\alpha\rho_0>4\pi$ \\
exterior mass fraction & $f_{\rm ext}\simeq0.1995$ at the $\alpha=0$ anchor \\
effective sound-speed surrogates & $c_{s,r}^2=-0.2$, $c_{s,t}^2=0.15$ \\
\bottomrule
\end{tabular}
\end{table}

Figure~\ref{fig:admissibility_wedge} makes the admissible wedge explicit. For the anchored branch $(w_r,w_t)=(-0.2,0.15)$ one also has $q=-0.9$, so the denominator $4\pi-\alpha q\rho_0$ in Eq.~\eqref{eq:modbalance} is bounded away from zero for $\alpha\ge0$ whenever $\rho_0\ge0$. Moreover, the closure-level combinations entering the null, weak, dominant, and strong energy conditions are all positive multiples of $\rho_0$; Table~\ref{tab:closure_audit} collects these diagnostic combinations together with the effective exterior-mass fraction inferred from the anchored background. The same table also shows why the constant-$(w_r,w_t)$ prescription should still be interpreted as an effective closure rather than a microphysical fluid model: the formal radial sound-speed surrogate $c_{s,r}^2=w_r$ is negative even though the branch is regular and hyperbolic in the odd sector.

As a representative admissible case, we use $w_r=-0.2$ and $w_t=0.15$, for which Eqs.~\eqref{eq:s_index} and \eqref{eq:ninfty} give $s=2$ and $n_\infty=3.5$. Appendix~B presents an independent outward integration for this case at $\alpha=0$ and confirms the expected $r^{-3.5}$ decay numerically. By contrast, the positive-$w_r$ family $w_r=+0.2$, $w_t=0.15$ yields the formal horizon exponent $s=-3$ and is therefore used below only as a comparison branch rather than as a self-consistent astrophysical background family.

\subsection{Axial perturbations and the compactified QNM system}
\label{subsec:oddsector}

In Regge--Wheeler gauge, the odd-parity metric perturbations are
\begin{equation}
h_{t\phi}=h_0(r)\e^{-i\omega t}S_\ell(\theta),
\qquad
h_{r\phi}=h_1(r)\e^{-i\omega t}S_\ell(\theta),
\end{equation}
with $S_\ell(\theta)=\sin\theta\,\partial_\theta P_\ell(\cos\theta)$. The axial fluid amplitude is written as
\begin{equation}
\delta u_\phi=r^2\varpi(r)\e^{-i\omega t}S_\ell(\theta).
\end{equation}
Odd parity does not perturb the scalar fluid variables,
\begin{equation}
\delta\rho=\delta p_r=\delta p_t=0,
\end{equation}
so the only nonvanishing perturbed stress-tensor components are
\begin{align}
\delta T_{t\phi}&=\Bigl[-(\rho_0+p_{t0})r^2\varpi+p_{t0}h_0\Bigr]\e^{-i\omega t}S_\ell,
\\
\delta T_{r\phi}&=p_{t0}h_1\e^{-i\omega t}S_\ell.
\end{align}
A key simplification is that
\begin{equation*}
\delta T=0,
\qquad
\delta f_T=0.
\end{equation*}
Even so, the $f_T$ sector does not disappear altogether. As shown in Appendix~A, the variation of $f_T(T_{\mu\nu}+\Theta_{\mu\nu})$ survives through $2\alpha T_0(\delta T_{\mu\nu}+\delta\Theta_{\mu\nu})$.

The axial matter equation follows from the modified divergence law, not from the GR conservation equation. The resulting linearized $\phi$-component is
\begin{equation}
-i\omega(\rho_0+p_{t0})\varpi+\Xi_0(r)h_1=0,
\end{equation}
where
\begin{equation}
\Xi_0(r)=\frac{p_{t0}-p_{r0}}{r}+(\rho_0+p_{t0})'-\frac{2\alpha p_{t0}T_0'}{8\pi-2\alpha T_0}.
\end{equation}
For $\omega\neq0$,
\begin{equation}
\varpi=\frac{\Xi_0(r)}{i\omega(\rho_0+p_{t0})}\,h_1.
\label{eq:varpi_elim}
\end{equation}
Substituting Eq.~\eqref{eq:varpi_elim} into the axial Einstein equations yields the coupled system
\begin{align}
h_0''+A_1h_0'+A_2h_0+A_3h_1'+A_4h_1&=0,
\label{eq:h0eq_main}
\\
h_1''+B_1h_1'+B_2h_1+B_3h_0'+B_4h_0&=0,
\label{eq:h1eq_main}
\end{align}
with coefficients
\begin{equation}
\begin{gathered}
A_1=\frac{2}{r}-\nu_0'+\mu_0',
\qquad
A_2=-e^{2(\nu_0-\mu_0)}\omega^2+e^{2\nu_0}(R_0+\alpha T_0^2+16\pi p_{t0}),
\\
A_3=i\omega,
\qquad
A_4=i\omega A_1-\frac{2e^{2\nu_0}r^2(8\pi-2\alpha T_0)}{i\omega}\,\Xi_0(r),
\\
B_1=\frac{2}{r}+\mu_0'-\nu_0',
\\
B_2=-e^{2(\nu_0-\mu_0)}\omega^2+\frac{e^{2\nu_0}}{r^2}\bigl(2-\ell(\ell+1)\bigr)+e^{2\nu_0}(R_0+\alpha T_0^2+16\pi p_{t0}),
\\
B_3=i\omega,
\qquad
B_4=-i\omega\nu_0'.
\end{gathered}
\label{eq:ABcoeffs_main}
\end{equation}
The explicit derivation is given in Appendix~A.

A complementary rewriting follows directly from Eq.~\eqref{eq:einsteinlike}. Defining
\begin{equation}
T_{\mu\nu}^{\rm eff}=c(T)\,T_{\mu\nu}+d(T,p_t)\,g_{\mu\nu},
\qquad
c(T)\equiv 1-\frac{\alpha T}{4\pi},
\qquad
d(T,p_t)\equiv \frac{\alpha}{8\pi}\left(2p_tT+\frac{1}{2}T^2\right),
\label{eq:Teff_main}
\end{equation}
the field equations become exactly
\begin{equation}
G_{\mu\nu}=8\pi T_{\mu\nu}^{\rm eff}.
\label{eq:einstein_eff_main}
\end{equation}
In odd parity the scalar fluid variables do not fluctuate, so $\delta\rho=\delta p_r=\delta p_t=\delta T=0$ and hence $\delta c=\delta d=0$. The odd sector is then identical to the odd-parity problem of Einstein gravity on a static background sourced by the frozen effective anisotropic fluid $T_{\mu\nu}^{\rm eff}$.

For a static line element written as
\begin{equation}
\dd s^2=-A(r)\dd t^2+\frac{\dd r^2}{B(r)}+C(r)\dd\Omega_2^2,
\end{equation}
the odd-parity sector can then be expressed in terms of a gauge-invariant master function $\Psi$ obeying \citep{MartelPoisson2005,FengPeng2024}
\begin{equation}
\frac{\dd^2\Psi}{\dd r_*^2}+\bigl[\omega^2-V_{\rm ax}(r)\bigr]\Psi=0,
\qquad
\frac{\dd r_*}{\dd r}=\frac{1}{\sqrt{AB}},
\label{eq:master_general_main}
\end{equation}
with
\begin{equation}
V_{\rm ax}(r)=\frac{[\ell(\ell+1)-2]A}{C}
+\sqrt{ABC}\,\frac{\dd}{\dd r}\left[\sqrt{AB}\,\frac{\dd}{\dd r}\left(C^{-1/2}\right)\right].
\label{eq:Vax_general_main}
\end{equation}
For the metric \eqref{eq:Nsigma_metric} one has
\begin{equation}
A=N\sigma^2,
\qquad
B=N,
\qquad
C=r^2,
\qquad
\Psi=N\sigma h_1,
\end{equation}
and therefore
\begin{equation}
V_{\rm ax}(r)=\frac{[\ell(\ell+1)-2]N\sigma^2}{r^2}
-\frac{N\sigma}{r}\frac{\dd(N\sigma)}{\dd r}
+\frac{2(N\sigma)^2}{r^2}.
\label{eq:Vax_ours_main}
\end{equation}
This exact master equation provides the production equation used below for the admissible negative-$w_r$ branch. The unreduced two-field system \eqref{eq:h0eq_main}--\eqref{eq:h1eq_main} remains the natural benchmark problem for the coupled PINN and for the positive-$w_r$ comparison branch.

For the unreduced coupled formulation used by the PINN benchmark, we compactify the exterior region with
\begin{equation}
x=\frac{r_H}{r},
\qquad
x\in[0,1],
\end{equation}
where $x=1$ is the horizon and $x=0$ is infinity. Let $r_*$ be the tortoise coordinate, $\dd r_*/\dd r=e^{\nu_0-\mu_0}$. Then the QNM conditions are purely ingoing at the horizon and purely outgoing at infinity. We factor these universal asymptotics as
\begin{equation}
h_a(x)=\mathcal{F}(x,\Omega)g_a(x),
\qquad a=0,1,
\end{equation}
with $\Omega=r_H\omega$ and
\begin{equation}
\mathcal{F}(x,\Omega)=(1-x)^{-i\gamma_H\Omega}x^{-2i\eta_M\Omega}\e^{i\Omega/x},
\qquad
\gamma_H\equiv\frac{1}{2\kappa_H r_H},
\qquad
\eta_M\equiv\frac{M}{r_H}.
\end{equation}
The regular functions $g_0$ and $g_1$ satisfy a compactified system of the schematic form
\begin{equation}
\mathcal{R}_a[g_0,g_1;\Omega](x)=\sum_{b=0}^{1}\sum_{n=0}^{2}\sum_m C_{nm}^{(ab)}(x)(i\Omega)^m\frac{\dd^n g_b}{\dd x^n}=0,
\qquad a=0,1,
\label{eq:compactified_operator}
\end{equation}
where the coefficient arrays are built directly from the background profiles and their derivatives. The Schwarzschild limit is recovered when $\rho_0=p_{r0}=p_{t0}=T_0=0$, in which case the system collapses to the usual Regge--Wheeler problem.

\section{Numerical strategy, validation hierarchy, and uncertainty model}
\label{sec:numerics}

\subsection{Numerical scope and validation strategy}
\label{subsec:numerical_scope}

We use two complementary numerical routes. The coupled PINN targets the unreduced two-field odd-parity eigenproblem and is employed here for the Schwarzschild benchmark and for the positive-$w_r$ comparison branch. For the physically admissible negative-$w_r$ family, the quoted frequencies instead come from an exact solve of the gauge-invariant axial master equation derived in Sec.~\ref{subsec:oddsector}. The coupled PINN and the compactified two-field Chebyshev formulation remain benchmark tools for the unreduced gauge-fixed problem, while the admissible-branch frequencies reported below come from the exact master-equation calculation.

\subsection{Coupled PINN architecture and exact horizon regularity}
\label{subsec:pinn}

The PINN learns the regularized fields $g_0$ and $g_1$ rather than $h_0$ and $h_1$ themselves. Its input is the single real coordinate $x\in[0,1]$. A shared encoder with four hidden layers of width 256 and GELU activations feeds two decoders, each with two hidden layers of width 128, which return the real and imaginary parts of the raw complex functions $G_0(x)$ and $G_1(x)$. The complex QNM frequency is represented by two trainable scalars, $\Omega_R$ and $\Omega_I$.

\begin{figure}[t]
\centering
\includegraphics[width=0.96\columnwidth]{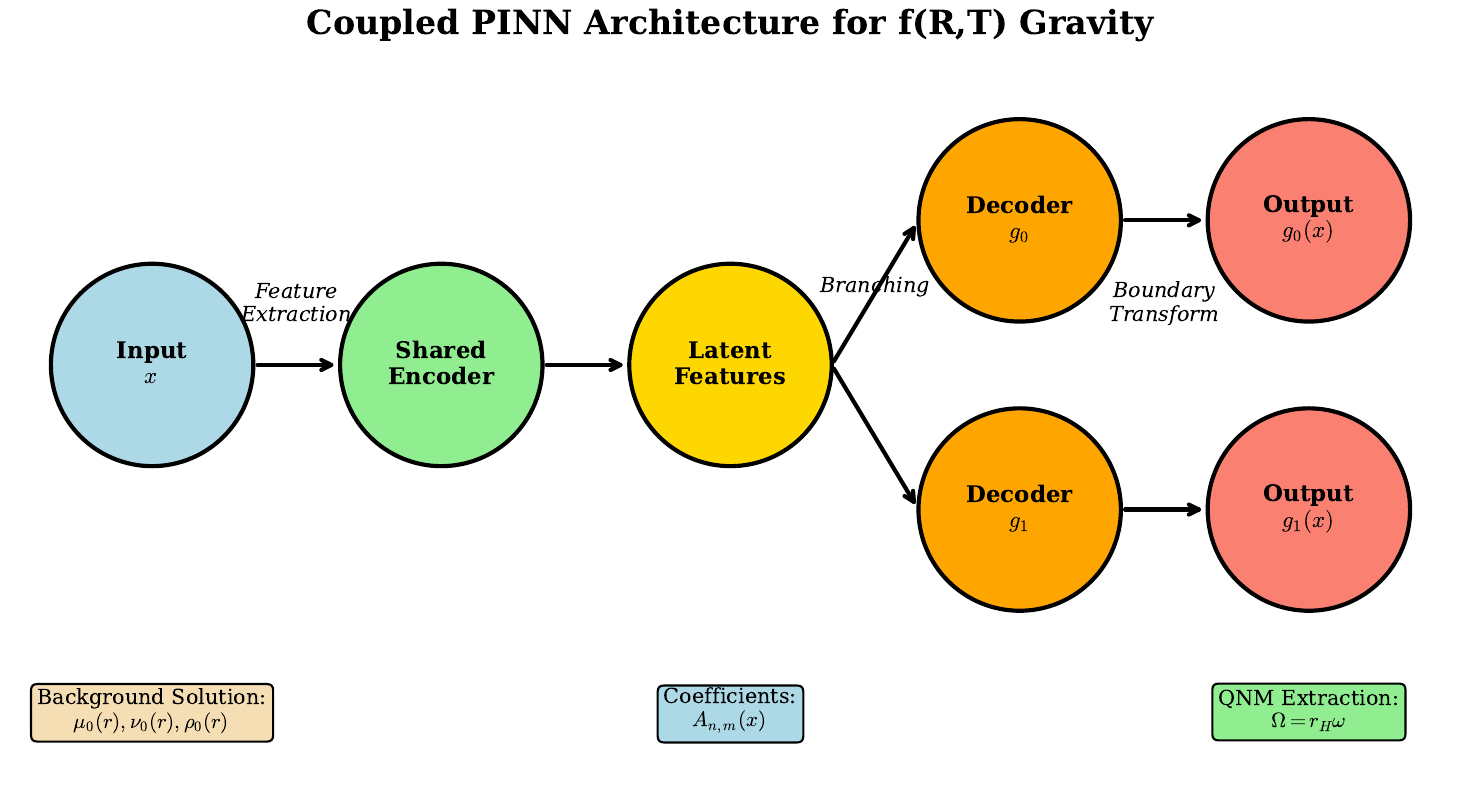}
\caption{Coupled PINN architecture for the compactified two-field axial eigenproblem. The shared encoder extracts common radial features, the two decoder branches generate field-specific latent outputs, and the boundary transform enforces Eq.~\eqref{eq:horizon_ansatz} to produce the regular fields $g_0$ and $g_1$. The physical perturbations are reconstructed as $h_a=\mathcal{F}g_a$.}
\label{fig:architecture}
\end{figure}

Because the QNM problem is homogeneous, one overall normalization must be fixed. Writing the near-horizon expansion as
\begin{equation}
g_a(x)=g_{a,H}+g_{a,1}(1-x)+O\!\bigl((1-x)^2\bigr),
\end{equation}
substitution into the regularized equations yields a linear relation
\begin{equation}
\mathcal{H}_{00}(\Omega)g_{0,H}+\mathcal{H}_{01}(\Omega)g_{1,H}=0.
\end{equation}
Defining
\begin{equation}
\Lambda_H(\Omega)\equiv-\frac{\mathcal{H}_{00}(\Omega)}{\mathcal{H}_{01}(\Omega)},
\end{equation}
we impose exact horizon regularity through
\begin{equation}
\begin{aligned}
g_0(x)&=1+(1-x)G_0(x),
\\
g_1(x)&=\Lambda_H(\Omega)+(1-x)G_1(x).
\end{aligned}
\label{eq:horizon_ansatz}
\end{equation}
This removes the need for a separate horizon penalty. No asymptotic penalty is required because the QNM behavior has already been factored into $\mathcal{F}(x,\Omega)$.

For the admissible branch, the quoted frequencies come directly from the exact gauge-invariant axial master equation on the anchored backgrounds, with Schwarzschild benchmarks and spectral-resolution checks described below. A full retrained coupled-PINN survey on the same branch would still be a useful cross-check, but the axial conclusions drawn here do not depend on it.

\subsection{Loss function and optimization protocol}
\label{subsec:loss}

At a collocation point $x_i$, let $\mathcal{R}_0(x_i)$ and $\mathcal{R}_1(x_i)$ denote the two complex residuals of the compactified system. We use automatic differentiation for the derivatives of $g_0$ and $g_1$ and the complex norm $|z|^2=(\Re z)^2+(\Im z)^2$. The equation-residual losses are
\begin{equation}
L_{\eq}^{(0)}=\frac{1}{N}\sum_{i=1}^{N}|\mathcal{R}_0(x_i)|^2,
\qquad
L_{\eq}^{(1)}=\frac{1}{N}\sum_{i=1}^{N}|\mathcal{R}_1(x_i)|^2.
\end{equation}
We also retain a coupling-sensitive regularizer,
\begin{equation}
L_{\coup}=\frac{1}{N}\sum_{i=1}^{N}\left|\frac{\partial\mathcal{R}_0}{\partial g_0}\frac{\partial\mathcal{R}_1}{\partial g_1}-\frac{\partial\mathcal{R}_0}{\partial g_1}\frac{\partial\mathcal{R}_1}{\partial g_0}\right|_{x_i}^2,
\end{equation}
but we now interpret it explicitly as an empirical numerical stabilizer rather than a derived physical condition. To suppress spurious short-wavelength oscillations in the learned regular functions we also include
\begin{equation}
L_{\sm}=\frac{1}{N}\sum_{i=1}^{N}\left(|g_0''(x_i)|^2+|g_1''(x_i)|^2\right).
\end{equation}
The total loss is
\begin{equation}
L_{\tot}=\lambda_0L_{\eq}^{(0)}+\lambda_1L_{\eq}^{(1)}+\lambda_cL_{\coup}+\lambda_sL_{\sm},
\label{eq:Ltot_mainloss}
\end{equation}
with initial weights $\lambda_0=\lambda_1=1$, $\lambda_c=10^{-1}$, and $\lambda_s=10^{-4}$. During the final stage of training the relative weights are updated adaptively to keep the four contributions comparable in scale.

Optimization is performed in three stages with Adam \citep{KingmaBa2015Adam}: a warm start with the off-diagonal couplings temporarily suppressed, a coupling ramp in which the full operator is restored gradually, and a final full-coupling stage with lower learning rate and cosine annealing. A typical production schedule is 1000 epochs for the warm start, 2000 for the coupling ramp, and 3000 for the final stage. Transfer learning is used along smooth one-parameter families so that the converged weights and frequency from one point initialize the next. The collocation set is nonuniform and concentrated toward both boundaries; the production value is $N=2000$ collocation points.

\subsection{Independent matrix method, hyperbolicity screen, and validation tiers}
\label{subsec:validation_tiers}

Rather than treating all nonzero-$\alpha$ points on the same footing, we assign each result to one of three validation tiers.
\begin{enumerate}
\item \textit{Externally validated}: the point satisfies the hyperbolicity criterion $\minH>10^{-4}$, the final loss obeys $L_{\tot}<0.1$, and an independent Chebyshev spectral solve agrees with the PINN result within the combined numerical uncertainty.
\item \textit{Internally validated}: the point satisfies $\minH>10^{-4}$ and $L_{\tot}<0.1$, and it is stable under seed variation, collocation refinement, architecture variation, and smooth continuation, but no external spectral solve is currently available.
\item \textit{Provisional or marginal}: the point is too close to the hyperbolicity boundary or the loss threshold for a strong quantitative claim. Such points are retained only to illustrate the approach to breakdown.
\end{enumerate}

For the positive-$w_r$ comparison branch, the independent spectral method discretizes the same compactified two-field operator on a Chebyshev--Gauss--Lobatto grid and solves the resulting matrix eigenproblem. After collocation, the problem takes a polynomial form in $z\equiv i\Omega$,
\begin{equation}
\bigl[\mathbf{A}_0+z\mathbf{A}_1+z^2\mathbf{A}_2\bigr]\mathbf{v}=0,
\label{eq:quadratic_evp_main}
\end{equation}
which can be linearized to a generalized eigenvalue problem as described explicitly in Appendix~D. This deterministic formulation provides the independent spectral cross-check used in the low-coupling regime of the raw coupled system. In practice, direct external comparison is reliable through $\alpha/M^2\lesssim0.1$, where the compactified operator remains well conditioned. At stronger coupling we therefore classify the comparison-branch results by validation tier rather than claim independent spectral confirmation. The admissible branch, by contrast, is quoted from the exact master-equation solve and its own spectral-resolution envelope.

Physical admissibility is determined from the unreduced time-domain odd system, not from the reduced frequency-domain pair. Before any PINN solve is attempted we collect the coefficients of the principal derivatives into the schematic form
\begin{equation}
\mathbf{K}(r)\,\partial_t^2\mathbf{q}+2\mathbf{M}(r)\,\partial_t\partial_r\mathbf{q}-\mathbf{P}(r)\,\partial_r^2\mathbf{q}+\cdots=0,
\label{eq:principal_form_main}
\end{equation}
where $\mathbf{q}$ denotes the unreduced odd variables. Inserting $\mathbf{q}\propto\exp[-i\omega t+ikr]$ gives the principal symbol
\begin{equation}
\boldsymbol{\Pi}(\omega,k;r)=-\omega^2\mathbf{K}-2\omega k\mathbf{M}+k^2\mathbf{P}.
\end{equation}
The characteristic phase speeds are the roots of
\begin{equation}
\det\!\bigl[c^2\mathbf{K}+2c\mathbf{M}-\mathbf{P}\bigr]=0,
\qquad c\equiv\omega/k.
\label{eq:charpoly_main}
\end{equation}
At each radius we sort the real roots and define $c_-^2(r)$ as the smallest physical branch. The background diagnostic used throughout the paper is
\begin{equation}
\minH(\alpha,w_r,w_t)\equiv\min_{x\in[0,1]}c_-^2(x).
\label{eq:Hmin_def_main}
\end{equation}
Hyperbolicity requires $\minH>0$. In practice we adopt the conservative stop criterion $\minH>10^{-4}$ to avoid grazing the boundary where roundoff and branch reordering become troublesome. Appendix~D gives the explicit extraction algorithm and the strong-shift validation protocol.

\subsection{Uncertainty budget for the admissible branch}
\label{subsec:uncertainty}

For the admissible branch, the quoted uncertainty is purely numerical and comes from the exact master-equation solve. Let $\Omega_{N_b,N_c}$ denote the frequency extracted on a given spectral grid and let $\Omega_{40,80}$ be the finest-grid value retained as the central estimate. Across the production grids $(N_b,N_c)=(28,56)$, $(32,64)$, $(36,72)$, and $(40,80)$ we assign
\begin{equation}
\sigma_{\Re(\Omega)}=\max_i\bigl|\Re(\Omega_i)-\Re(\Omega_{40,80})\bigr|,
\qquad
\sigma_{-\Im(\Omega)}=\max_i\bigl|\Im(\Omega_i)-\Im(\Omega_{40,80})\bigr|,
\label{eq:unc_budget_adm}
\end{equation}
which at the $\alpha=0$ anchor gives
\begin{equation}
\sigma_{\Re(\Omega)}=2.98\times10^{-5},
\qquad
\sigma_{-\Im(\Omega)}=2.23\times10^{-5}.
\label{eq:unc_values_adm}
\end{equation}
The corresponding mass-normalized uncertainties are obtained by multiplying by $M$, giving $\sigma_{\Re(M\omega)}\simeq1.86\times10^{-5}$ and $\sigma_{-\Im(M\omega)}\simeq1.39\times10^{-5}$ at the anchor. For the adopted asymptotic fit itself, the least-squares intercept uncertainty in $M$ is much smaller: at both the anchor and the $\beta=0.3$ endpoint we find $\delta M/M\simeq4.7\times10^{-8}$ for the tail fit used in the production pipeline. Even when the fit window is varied across the asymptotic region, the extracted mass changes only at the $\sim1.5\times10^{-6}$ level in relative terms, so the induced contribution to $M\omega$ stays below $7\times10^{-7}$ and remains negligible compared with the spectral-resolution envelope. We therefore scale the horizon-normalized uncertainties by the central value of $M$ and adopt this envelope conservatively for every row of Table~\ref{tab:negativewr_scan}, without transferring seed scatter or architecture variation from the positive-$w_r$ comparison branch to the admissible branch.

\subsection{Internal PINN scatter on the comparison branch}
\label{subsec:comparison_uncertainty}

Seed-to-seed PINN scatter and architecture sensitivity are reported only for the positive-$w_r$ comparison branch, where the full coupled-PINN campaign was performed. Those numbers are useful for method characterization but are not quoted as the uncertainty model for the admissible-branch frequencies. For interpretation of the comparison calculation we adopt the conservative fractional envelope
\begin{equation}
\epsilon_{\mathrm{comp}}^2=\epsilon_{\mathrm{seed}}^2+\epsilon_{\mathrm{coll}}^2+\epsilon_{\mathrm{arch}}^2+\epsilon_{\mathrm{spec}}^2+\epsilon_{\mathrm{loss}}^2,
\label{eq:unc_budget}
\end{equation}
where $\epsilon_{\mathrm{spec}}$ is included only when the independent Chebyshev check exists and $\epsilon_{\mathrm{loss}}$ is an empirical monotone proxy tied to the final residual level. The machine-readable file \texttt{uncertainty\_budget.csv} lists the adopted pointwise envelopes for the positive-$w_r$ comparison scan. They rise from $0.415\%$ at $\alpha/M^2=0$ to $0.457\%$ at $0.1$, $0.930\%$ at $0.25$, $1.34\%$ at $0.30$, and $1.78\%$ at the marginal point $0.325$.

Representative components are listed in Table~\ref{tab:uncertainty}. Two features are especially important: in the externally validated regime the architecture spread already exceeds the seed scatter, and near the loss threshold the residual-based term becomes significant. These comparison-branch numbers should not be read as the uncertainty model for the admissible branch.

\begin{table*}[t]
\centering
\small
\setlength{\tabcolsep}{6pt}
\renewcommand{\arraystretch}{1.08}
\caption{Representative PINN uncertainty budget for the positive-$w_r$ comparison branch. Percentage entries are fractional uncertainties. The seed contribution is the larger of the relative uncertainties in $\Re(\Omega)$ and $-\Im(\Omega)$ quoted by the comparison scan. The loss proxy is normalized so that $L_{\tot}=0.1$ corresponds to a $1\%$ solver-bias envelope.}
\label{tab:uncertainty}
\begin{tabular*}{\textwidth}{@{\extracolsep{\fill}}lcc@{}}
\toprule
Source & $\alpha/M^2=0.10$ & $\alpha/M^2=0.30$ \\
\midrule
Seed scatter & $0.163\%$ & $0.942\%$ \\
Collocation refinement & $0.040\%$ & \shortstack[c]{$0.040\%$\\(carried conservatively)} \\
Architecture and activation variation & $0.400\%$ & $0.400\%$ \\
External Chebyshev comparison & $0.100\%$ & unavailable \\
Residual-level proxy & $0.102\%$ & $0.870\%$ \\
Adopted conservative envelope & $0.457\%$ & $1.34\%$ \\
\bottomrule
\end{tabular*}
\end{table*}

\section{Results}
\label{sec:results}

\subsection{Background consistency audit}
\label{subsec:consistency_audit}

The background analysis fixes the physically admissible sector. For the constant-$(w_r,w_t)$ closure, nontrivial asymptotically flat matter-supported branches satisfy the horizon and asymptotic constraints derived in Section~\ref{subsec:background}, namely
\begin{equation}
-\frac{1}{3}\le w_r<0,
\qquad
n_\infty=\frac{2(w_r-w_t)}{w_r}>3,
\qquad
w_t>-\frac{w_r}{2}.
\label{eq:admissible_sector_main}
\end{equation}
Equation~\eqref{eq:admissible_sector_main} defines the wedge shown in Figure~\ref{fig:admissibility_wedge}. The reference family $w_r=0.2$ and $w_t=0.15$ violates the first condition, giving the formal horizon exponent $s=-3$, and therefore cannot represent a regular asymptotically flat matter-supported background of the stated model. By contrast, the anchored branch $(w_r,w_t)=(-0.2,0.15)$ also passes the closure-level audit summarized in Table~\ref{tab:closure_audit}: for $\rho_0>0$ the diagnostic NEC/WEC/DEC/SEC combinations are positive, the modified-balance denominator is safe for $\alpha\ge0$, and the inferred exterior effective-matter fraction is about $20\%$. The positive-$w_r$ results shown below are included only as a comparison scan near the admissibility boundary, not as physical predictions.

Appendix~B confirms that the admissible sector is nonempty through an independent outward integration for the representative case $(w_r,w_t,\alpha)=(-0.2,0.15,0)$. The analytic prediction is $s=2$ at the horizon and $n_\infty=3.5$ at large radius; the numerical fit gives $n_{\rm fit}=3.497$ and $M/r_H=0.624628$, in excellent agreement with that scaling. This background anchors the exact admissible-branch calculation used in the next subsection.

\subsection{Validated calculation on the admissible negative-\texorpdfstring{$w_r$}{w_r} branch}
\label{subsec:negativewr_branch}

We now turn to the admissible family singled out by the background analysis. Throughout this section the branch is anchored at
\begin{equation}
(w_r,w_t)=(-0.2,0.15),
\qquad
\rho_0(r)=\rho_s(r-r_H)^2+\cdots,
\qquad
\rho_s=0.01835620783535,
\label{eq:negative_branch_anchor}
\end{equation}
with $r_H=1$ and with the shooting amplitude fixed so that the $\alpha=0$ solution reproduces the Appendix~B mass ratio
\begin{equation}
\left.\frac{M}{r_H}\right|_{\alpha=0}=0.6246281367.
\label{eq:negative_branch_mass_anchor}
\end{equation}
The quoted value of $\rho_s$ is normalization dependent and is given only to make the anchored branch reproducible. For each target $\beta\equiv\alpha/M^2$, we solve the fixed-point relation
\begin{equation}
\frac{\alpha}{r_H^2}=\beta\left[\frac{M(\alpha)}{r_H}\right]^2
\label{eq:beta_fixedpoint}
\end{equation}
self-consistently and continue the same anchored branch up to $\beta=0.3$. Across this scan $M/r_H$ changes only from $0.6246281367$ to $0.6246281642$, while $\gamma_H$ changes only from $1.0245412638$ to $1.0245412829$. Hyperbolicity is likewise clean on this admissible family: Appendix~\ref{app:master_exact} reduces the odd sector to a single wave equation with principal part $-\partial_t^2+\partial_{r_*}^2$, so in that normalization the corresponding characteristic-speed diagnostic stays at $\minH=1$ up to interpolation-level numerical noise throughout the anchored scan.

\begin{figure}[t]
\centering
\includegraphics[width=0.98\columnwidth]{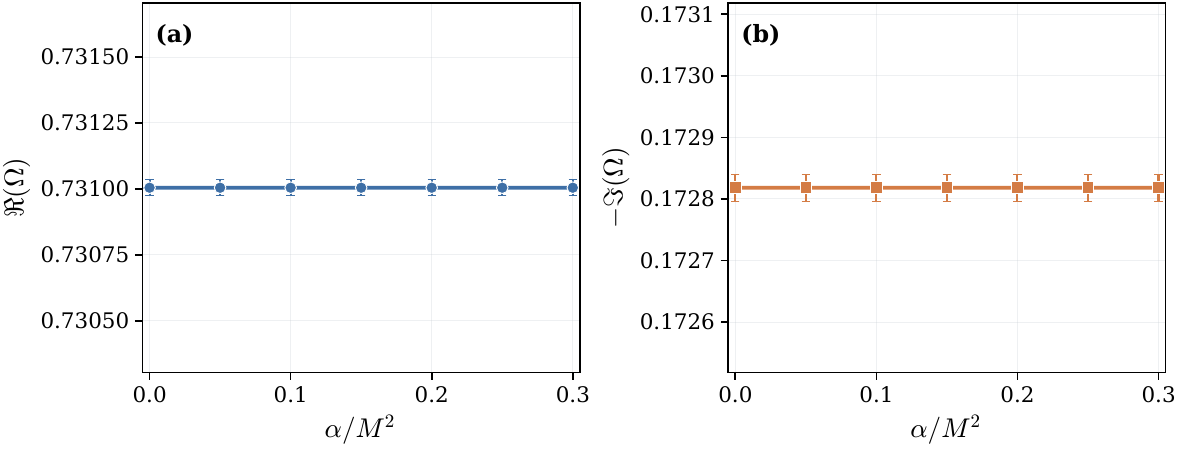}
\caption{Production scan on the anchored admissible negative-$w_r$ branch, $(w_r,w_t)=(-0.2,0.15)$ with $\rho_s=0.01835620783535$. Left: $\Re(\Omega)$. Right: $-\Im(\Omega)$. Points show the finest-grid frequencies from the exact axial master equation, dashed horizontal lines mark the $\alpha=0$ baselines, and error bars show the conservative spectral envelope from inter-grid spread. The scan is effectively flat: the pointwise central values vary by less than $4\times10^{-8}$ across $0\le\alpha/M^2\le0.3$, far below the quoted uncertainty envelope.}
\label{fig:negativewr_scan}
\end{figure}

\begin{figure}[t]
\centering
\includegraphics[width=0.98\columnwidth]{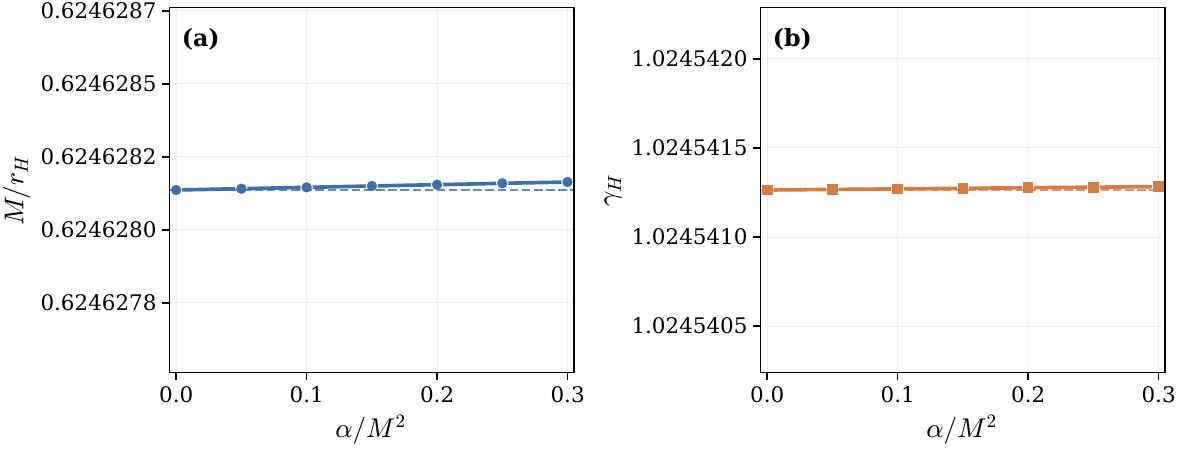}
\caption{Background quantities along the anchored admissible negative-$w_r$ branch. Left: $M/r_H$. Right: $\gamma_H$. Dashed horizontal lines mark the $\alpha=0$ baselines. Both remain essentially constant across the full scan, which explains the negligible direct $\alpha$-dependence of the ringdown spectrum.}
\label{fig:negativewr_background}
\end{figure}

We then solve this family with the exact gauge-invariant axial master equation of Sec.~\ref{subsec:oddsector}. The production scan uses the spectral grids described in Sec.~\ref{subsec:uncertainty}; the finest-grid value is taken as the central frequency, and the conservative envelope is set by the largest inter-grid deviation. The anchored branch yields the baseline fundamental mode
\begin{align}
\Omega_{\rm adm}&=0.7310046\pm0.0000298-(0.1728182\pm0.0000223)i,\\
M\omega_{\rm adm}&=0.4566061\pm0.0000186-(0.1079471\pm0.0000139)i.
\label{eq:negative_branch_baseline_mode}
\end{align}
Relative to Schwarzschild, the $r_H$-normalized shifts are $-2.19\%$ in $\Re(\Omega)$ and $-2.87\%$ in $-\Im(\Omega)$, while the mass-normalized shifts are $+22.19\%$ and $+21.34\%$ because the admissible branch has $M/r_H>1/2$. Because $(M/r_H)_{\rm adm}=0.6246281367$, the anchored branch also carries an effective exterior mass fraction
\begin{equation}
f_{\rm ext}\equiv 1-\frac{r_H/2}{M}\simeq 0.1995,
\label{eq:fext_def}
\end{equation}
so the $\sim22\%$ shift in $M\omega$ is associated with an $\sim20\%$ exterior contribution rather than with a resolved direct change of the axial operator along the branch. Along the branch itself, the pointwise central values vary by less than $4\times10^{-8}$ across $0\le\beta\le0.3$, far below the conservative spectral envelope. Table~\ref{tab:negativewr_scan} and Figures~\ref{fig:negativewr_scan}--\ref{fig:negativewr_background} therefore show an axially stable admissible branch whose main spectral departure from Schwarzschild comes from the existence of the matter-supported exterior itself. The large monotonic strong-coupling shifts seen in the positive-$w_r$ comparison scan disappear once the admissibility conditions are enforced.

\begin{table*}[t]
\centering
\small
\setlength{\tabcolsep}{6pt}
\renewcommand{\arraystretch}{1.08}
\caption{Representative points from the exact production scan on the anchored admissible branch $(w_r,w_t)=(-0.2,0.15)$. Display precision is rounded for readability; the machine-readable tables retain full numerical precision. The quoted frequencies are the finest-grid values from the exact axial master equation, and every row carries the same conservative spectral envelope, $\sigma_{\Re\Omega}=2.98\times10^{-5}$, $\sigma_{-\Im\Omega}=2.23\times10^{-5}$, $\sigma_{\Re(M\omega)}\simeq1.86\times10^{-5}$, and $\sigma_{-\Im(M\omega)}\simeq1.39\times10^{-5}$.}
\label{tab:negativewr_scan}
\begin{tabular*}{\textwidth}{@{\extracolsep{\fill}}cccccc@{}}
\toprule
$\alpha/M^2$ & $M/r_H$ & $\Re(\Omega)$ & $-\Im(\Omega)$ & $\Re(M\omega)$ & $-\Im(M\omega)$ \\
\midrule
0.00 & 0.62462814 & 0.7310046 & 0.1728182 & 0.4566061 & 0.1079471 \\
0.10 & 0.62462815 & 0.7310046 & 0.1728182 & 0.4566061 & 0.1079471 \\
0.20 & 0.62462816 & 0.7310046 & 0.1728182 & 0.4566061 & 0.1079471 \\
0.30 & 0.62462816 & 0.7310046 & 0.1728182 & 0.4566061 & 0.1079471 \\
\bottomrule
\end{tabular*}
\end{table*}

\subsection{Mass-normalized spectra and ringdown diagnostics}
\label{subsec:obs_results}

\begin{figure}[t]
\centering
\includegraphics[width=0.98\columnwidth]{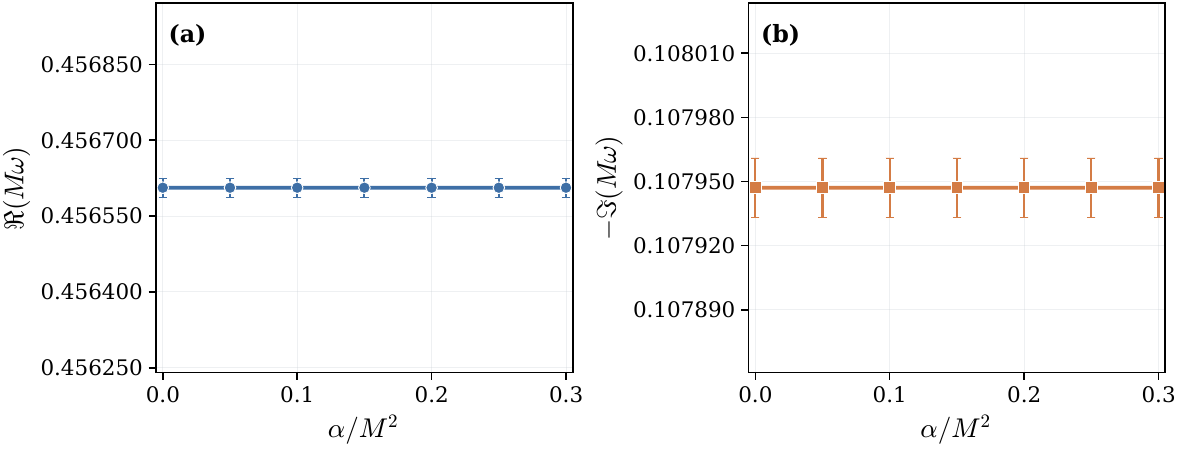}
\caption{Mass-normalized axial QNM frequencies on the anchored admissible negative-$w_r$ branch. Left: $\Re(M\omega)$. Right: $-\Im(M\omega)$. Dashed horizontal lines mark the $\alpha=0$ baselines, and error bars show the same conservative spectral envelope used in Table~\ref{tab:negativewr_scan}. The central values vary by less than $4\times10^{-8}$ across the scan, so the visible spread is entirely set by the conservative uncertainty estimate.}
\label{fig:negativewr_massnorm}
\end{figure}

Mass-normalized frequencies are the quantities most directly tied to fixed-mass ringdown observables. Figure~\ref{fig:negativewr_massnorm} shows that the admissible branch remains effectively flat in both $\Re(M\omega)$ and $-\Im(M\omega)$ across $0\le\alpha/M^2\le0.3$, mirroring the horizon-normalized scan. At $\alpha=0$ we find
\begin{equation}
\Re(M\omega)=0.4566061\pm0.0000186,
\qquad
-\Im(M\omega)=0.1079471\pm0.0000139.
\end{equation}
Relative to Schwarzschild, these correspond to shifts of $+22.19\%$ and $+21.34\%$. The net mass-normalized shift can be written as
\begin{equation}
\frac{(M\omega)_{\rm adm}}{(M\omega)_{\Schw}}
=
\frac{(M/r_H)_{\rm adm}}{1/2}\,
\frac{\Omega_{\rm adm}}{\Omega_{\Schw}},
\label{eq:branch_decomp}
\end{equation}
so the $24.93\%$ increase in $M/r_H$ is partly offset by the $2.19\%$ and $2.87\%$ reductions in $\Re(\Omega)$ and $-\Im(\Omega)$. The branch shift exceeds the conservative spectral envelope by more than three orders of magnitude, whereas the residual direct $\alpha$-dependence along the branch is still far below that same envelope.

At fixed source mass, the corresponding ringdown frequency $f_{220}$ and damping time $\tau_{220}$ follow
\begin{equation}
f_{220}=\frac{\Re(M\omega)}{2\pi (GM/c^3)},
\qquad
\tau_{220}=\frac{GM/c^3}{-\Im(M\omega)}.
\end{equation}
The admissible branch therefore raises $f_{220}$ by $22.19\%$ and shortens $\tau_{220}$ by $17.59\%$, while the quality factor
\begin{equation}
Q=\frac{\Re(M\omega)}{2[-\Im(M\omega)]}
\end{equation}
changes by only $0.70\%$ relative to Schwarzschild. Using the same one-parameter distinguishability scaling $\rho_{\rm rd}\sim1/\delta$ only as an order-of-magnitude guide, the branch-level offsets from Schwarzschild correspond to target ringdown SNRs of a few, whereas resolving the residual direct $\alpha$-driven drift along the anchored branch would require prohibitively large SNRs, $\rho_{\rm rd}\gtrsim10^8$. These numbers are not detector forecasts: waveform systematics, multimode fitting, and pseudospectral or non-normal effects would need a dedicated inference study before any observational claim could be made \citep{DestounisDuque2024}. Table~\ref{tab:admissible_mass_examples} lists representative fixed-mass values for stellar-mass and massive-black-hole sources.

\begin{table}[t]
\centering
\small
\setlength{\tabcolsep}{4.5pt}
\renewcommand{\arraystretch}{1.08}
\caption{Representative fixed-mass ringdown observables computed from the baseline admissible branch and the exact Schwarzschild $\ell=2$ fundamental mode. Frequencies use $f_{220}=\Re(M\omega)/[2\pi(GM/c^3)]$ and damping times use $\tau_{220}=(GM/c^3)/[-\Im(M\omega)]$. Because $f_{220}\propto M^{-1}$ and $\tau_{220}\propto M$, the same fractional shifts apply at any mass.}
\label{tab:admissible_mass_examples}
\begin{tabular*}{0.98\columnwidth}{@{\extracolsep{\fill}}ccccc@{}}
\toprule
$M/M_\odot$ & $f_{220}^{\Schw}$ [Hz] & $f_{220}^{\rm adm}$ [Hz] & $\tau_{220}^{\Schw}$ [s] & $\tau_{220}^{\rm adm}$ [s] \\
\midrule
30 & 402.5 & 491.8 & $1.661\times10^{-3}$ & $1.369\times10^{-3}$ \\
60 & 201.2 & 245.9 & $3.322\times10^{-3}$ & $2.738\times10^{-3}$ \\
$10^6$ & $1.207\times10^{-2}$ & $1.475\times10^{-2}$ & 55.37 & 45.63 \\
\bottomrule
\end{tabular*}
\end{table}

\begin{figure}[t]
\centering
\includegraphics[width=0.98\columnwidth]{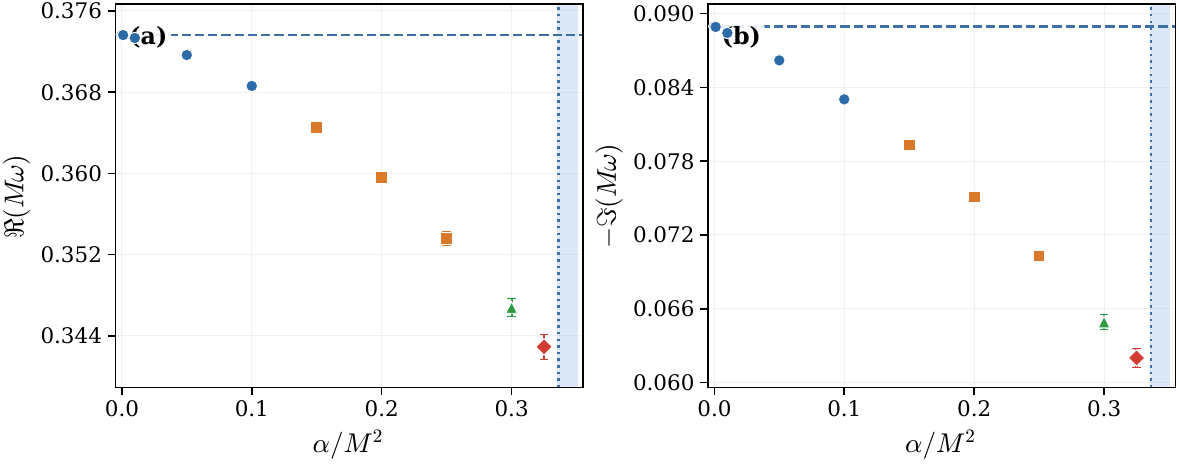}
\caption{Mass-normalized axial QNM frequencies for the positive-$w_r$ comparison branch. Left: $\Re(M\omega)$. Right: $-\Im(M\omega)$. Dashed horizontal lines mark the $\alpha=0$ values, while the dotted vertical line and shaded band indicate the same near-boundary regime shown in Figure~\ref{fig:alpha_scan}. The trend is monotonic in $\alpha/M^2$, but the figure is included only to illustrate the approach to the admissibility boundary because the underlying branch fails the asymptotic-regularity test.}
\label{fig:massnorm}
\end{figure}

For contrast, Figure~\ref{fig:massnorm} shows the monotonic mass-normalized spectrum on the positive-$w_r$ comparison branch as the admissibility boundary is approached. Those stronger-coupling trends remain useful as a numerical benchmark, but because the underlying backgrounds fail the regularity test they should not be interpreted as astrophysical predictions.

\subsection{Comparison with the positive-\texorpdfstring{$w_r$}{w_r} branch}
\label{subsec:baseline_results}

We keep the positive-$w_r$ family only as a numerical benchmark. Since it violates the regularity constraints derived above, it does not enter the physical interpretation. Its value is methodological: it tests the coupled-PINN solver and shows how apparently strong coupling-driven shifts can arise on backgrounds that fail the admissibility test.

The vacuum Schwarzschild limit provides the first reference check. Setting $\rho_0=p_{r0}=p_{t0}=T_0=0$ recovers the standard vacuum geometry and the Regge--Wheeler boundary-value problem. Our pipeline returns
\begin{equation}
\Omega_{\PINN}^{\Schw}=0.74730(5)-0.17792(3)i,
\end{equation}
in agreement with the exact Schwarzschild value $\Omega_{\Schw}=0.747343-0.177925i$ at the level of a few parts in $10^{-5}$. Appendix~C derives the first-order formula
\begin{equation}
\Omega_1=-\frac{\langle\chi_0,\mathcal{L}_1\mathbf{g}_0\rangle}{\langle\chi_0,(\partial_\Omega\mathcal{L}_0)\mathbf{g}_0\rangle},
\label{eq:omega1_formal_main}
\end{equation}
which is the quantity used to build the linearized reference curve. For the smallest nonzero couplings in the comparison scan, the discrepancy $|\Omega_{\PINN}-\Omega_{\mathrm{lin}}|$ scales approximately as $\alpha^2$, which is consistent with the perturbative expectation. Those small-$\alpha$ points, together with the Chebyshev cross-check up to $\alpha/M^2=0.1$, define the externally validated regime of the positive-$w_r$ comparison scan.

\begin{figure}[t]
\centering
\includegraphics[width=0.98\columnwidth]{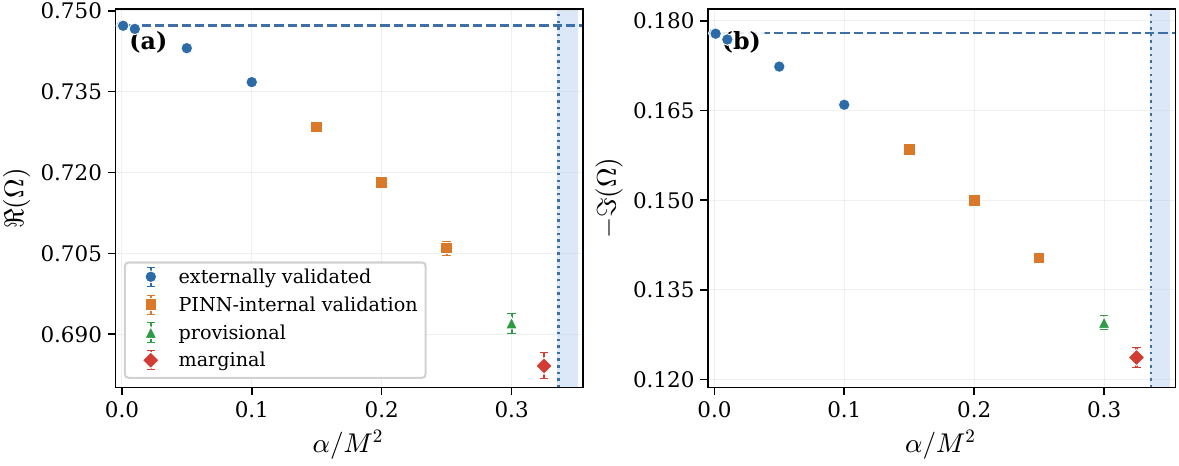}
\caption{Frequencies on the positive-$w_r$ comparison branch for $w_r=0.2$ and $w_t=0.15$. Left: $\Re(\Omega)$. Right: $-\Im(\Omega)$. Dashed horizontal lines show the Schwarzschild values. The dotted vertical line marks the interpolated hyperbolicity zero, $\alpha_{\mathrm{crit}}/M^2\approx0.336$, and the shaded band indicates the near-boundary regime where the linear problem becomes marginal or inadmissible. Circles denote externally validated points ($\alpha/M^2\le0.1$), squares denote points supported only by internal PINN validation, the triangle at $\alpha/M^2=0.30$ is provisional, and the diamond at $\alpha/M^2=0.325$ is marginal because it exceeds the practical loss threshold. Because this branch fails the regularity conditions derived in Section~\ref{subsec:background}, the figure should be read only as a numerical comparison of the trend toward the admissibility boundary.}
\label{fig:alpha_scan}
\end{figure}

Figure~\ref{fig:alpha_scan} shows the positive-$w_r$ comparison scan with each point labeled by validation tier. The dotted vertical line marks the linear interpolation of the zero of $\minH$ from the tabulated values at $\alpha/M^2=0.325$ and $0.35$: 
\begin{equation}
\alpha_{\mathrm{crit}}/M^2\approx0.336.
\label{eq:acrit_interp}
\end{equation}
The practical threshold $\minH=10^{-4}$ lies at essentially the same location, so we treat $\alpha/M^2=0.30$ as the last point worth emphasizing in the comparison scan and $\alpha/M^2=0.325$ only as a marginal reference.

The full pointwise frequency catalogue for this comparison branch is collected in Appendix~\ref{app:comparisoncatalogues} so that the main text can remain focused on the admissible branch.

Within this comparison scan, both the oscillation frequency $\Re(\Omega)$ and the damping-rate proxy $-\Im(\Omega)$ decrease monotonically as $\alpha/M^2$ increases. At $\alpha/M^2=0.10$ the relative shifts are about $-1.41\%$ and $-6.72\%$, while at the last provisional point, $\alpha/M^2=0.30$, they reach roughly $-7.39\%$ and $-27.2\%$. Those values are informative as numerical trends, but they are not final physical predictions because the underlying branch is not self-consistent.

\begin{figure}[t]
\centering
\includegraphics[width=0.98\columnwidth]{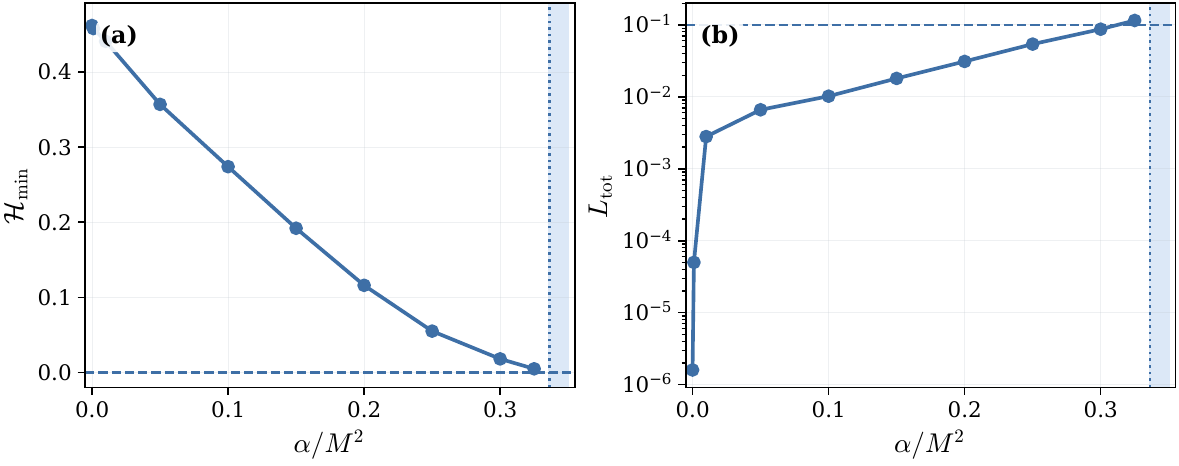}
\caption{Admissibility diagnostics for the positive-$w_r$ comparison branch. Left: the minimum characteristic-speed diagnostic $\minH$. Right: the final total loss on a logarithmic scale. The dashed horizontal lines mark $\minH=0$ and $L_{\tot}=0.1$, respectively. The dotted vertical line marks the interpolated hyperbolicity zero at $\alpha/M^2\approx0.336$, and the shaded band denotes the near-boundary regime where the extracted frequency is only marginally trustworthy.}
\label{fig:hyper_loss}
\end{figure}

Figure~\ref{fig:hyper_loss} separates the physical and numerical diagnostics. Both quantities move monotonically toward breakdown, but they carry different meanings: the rise in $L_{\tot}$ is a numerical warning sign, whereas the decrease of $\minH$ is a physical admissibility test derived from the principal symbol. We therefore treat the principal-symbol boundary as primary and the loss threshold as secondary.

\paragraph{Auxiliary anisotropy comparison.}
An auxiliary scan at fixed $\alpha/M^2=0.1$ is collected in Appendix~\ref{app:comparisoncatalogues}, together with the basis-dependent admixture diagnostic $\Ag$. Even within this inadmissible comparison family, the spectrum is much more sensitive to $w_t$ than to $w_r$, consistent with the coefficient structure in Eq.~\eqref{eq:ABcoeffs_main}. We leave those data in the appendix because they are numerically useful but do not affect the physical interpretation.

\section{Discussion}
\label{sec:discussion}

Two points stand out on the admissible branch. First, within the constant-$(w_r,w_t)$ closure, the axial spectrum is essentially insensitive to direct variation of $\alpha$ along the anchored family: over $0\le\alpha/M^2\le0.3$, the pointwise drift in $M\omega$ stays far below the conservative spectral envelope of the exact master-equation solve. Second, the branch itself is not close to Schwarzschild. Relative to the vacuum baseline, the anchored solution shifts $\Re(M\omega)$ and $-\Im(M\omega)$ by about $22\%$. Because $(M/r_H)_{\rm adm}=0.624628$, that shift tracks an effective exterior mass fraction of about $20\%$. In other words, the dominant axial signature comes from the presence of the matter-supported exterior, not from a directly resolved change in the trace coupling along the branch.

That conclusion rests on a direct solve of the exact axial master equation on self-consistent admissible backgrounds, benchmarked against Schwarzschild and tracked across the $\beta$ scan. The coupled PINN remains valuable because it solves the unreduced gauge-fixed problem directly and provides a useful benchmark on the comparison branch, while the admissible-branch spectrum itself is taken from the exact master equation rather than from a surrogate reduction or transferred PINN scatter. Within this effective closure, ``physical admissibility'' means regularity, asymptotic flatness, odd-sector hyperbolicity, positive diagnostic NEC/WEC/DEC/SEC combinations, and denominator safety in the modified balance law. At the same time, the negative formal surrogate $c_{s,r}^2=w_r$ makes it clear that the constant-$(w_r,w_t)$ prescription should be viewed as an effective anisotropic stress model, not as a literal microphysical fluid. That caveat is intrinsic to \fRT{} gravity, where different matter-sector completions can realize the same formal Lagrangian in different ways.

From an observational standpoint, the fixed-mass shifts in Table~\ref{tab:admissible_mass_examples} are large enough to make the branch interesting, but the fundamental axial mode by itself is not a clean discriminator of the trace-quadratic coupling. The quality factor differs from Schwarzschild by only $0.70\%$, so a Kerr source with slightly adjusted mass and spin can sit close to the same $(f_{220},\tau_{220})$ point when only one mode is used \citep{BertiWill2006,Berti2018}. A serious spectroscopy statement will therefore require multimode waveform modeling, detector response, parameter covariances, and possibly non-normal effects in black-hole perturbation problems \citep{DestounisDuque2024}. The polar sector remains the clearest next target. In odd parity $\delta T=0$, so the trace coupling enters only indirectly; in polar parity $\delta T\neq0$, direct trace-coupling terms appear already at linear order. On the anchored backgrounds, $\max_r |\alpha T_0|/(4\pi)\simeq2.0\times10^{-6}$ at the top of the scan, so any generic polar effect on this branch is more naturally expected at the $10^{-5}$ level unless it is enhanced by mode mixing or by a near-degeneracy of the polar operator. An independent time-domain extraction on the same backgrounds would be a useful next check.

\section{Conclusions}
\label{sec:conclusions}

We have identified the regular, asymptotically flat, hyperbolic odd-sector branch of the constant-$(w_r,w_t)$ closure in trace-quadratic $f(R,T)$ gravity and derived its axial perturbation problem. The admissible branch lies at negative $w_r$; the commonly used positive-$w_r$ family fails the regularity test and is useful only as a comparison branch. On admissible static backgrounds the odd sector reduces exactly to a single gauge-invariant master equation, which allows a direct solve of the axial spectrum.

For the branch anchored at $(w_r,w_t)=(-0.2,0.15)$, the baseline axial $\ell=2$ mode is
\begin{equation}
\Omega_{\rm adm}=0.7310046\pm0.0000298-(0.1728182\pm0.0000223)i,
\end{equation}
with
\begin{equation}
M\omega_{\rm adm}=0.4566061\pm0.0000186-(0.1079471\pm0.0000139)i.
\end{equation}
Relative to Schwarzschild, the branch shifts the mass-normalized oscillation frequency and damping rate by $+22.19\%$ and $+21.34\%$, respectively, while the direct drift with $\alpha$ along the same family remains numerically unresolved within the conservative spectral envelope over $0\le\alpha/M^2\le0.3$.

Within the constant-$(w_r,w_t)$ closure, axial ringdown therefore probes the existence of the admissible matter-supported branch much more readily than it probes the trace-quadratic coupling itself. The natural next steps are an independent time-domain extraction on the same backgrounds and a polar-sector analysis, where direct $\delta T$ effects can appear already at linear order.

More broadly, the analysis shows that an admissibility-first treatment can qualitatively reshape ringdown forecasts in matter-coupled modified gravity. Apparent strong-coupling trends on formal branches can disappear once regularity, asymptotic flatness, and odd-sector hyperbolicity are enforced. That lesson is likely to matter beyond the specific closure studied here.

\section*{Data availability}
A source bundle accompanies this manuscript. It contains the \LaTeX{} source, figure files, machine-readable tables, plotting scripts, the background-integration scripts, the exact master-equation code used for the admissible-branch scan and Schwarzschild benchmark, and the coupled-PINN comparison data products for the unreduced two-field problem. The machine-readable tables cover the admissible-branch production scan, fixed-mass examples, comparison-branch catalogues, and uncertainty budgets used throughout the paper. The complete bundle is available from the authors on request during review and will be deposited in a public Zenodo repository with a persistent DOI upon acceptance.

\clearpage
\appendix
\renewcommand{\theHequation}{\thesection\arabic{equation}}

\section{Derivation of the odd-parity perturbation equations in quadratic $f(R,T)$ gravity}

This appendix derives the axial perturbation equations used in the main text. In particular, it makes explicit how the odd-parity linearization is carried out in the quadratic trace-coupled model
\begin{equation*}
f(R,T)=R+\alpha T^2,
\end{equation*}
and how the matter-coupling term proportional to $f_T$ enters at linear order.

Throughout we use metric signature $(-,+,+,+)$, adopt standard curvature conventions as in Refs.~\cite{MTW1973,Wald1984Book,Weinberg1972Book}, and let primes denote derivatives with respect to $r$.

\subsection{Full field equations and matter-sector identities}

We begin from the $f(R,T)$ field equations
\begin{equation}
f_R R_{\mu\nu}-\frac{1}{2}f g_{\mu\nu}+(g_{\mu\nu}\Box-\nabla_\mu\nabla_\nu)f_R
=8\pi T_{\mu\nu}+f_T(T_{\mu\nu}+\Theta_{\mu\nu}),
\end{equation}
with
\begin{equation*}
f_R=\frac{\partial f}{\partial R}, 
\qquad 
f_T=\frac{\partial f}{\partial T}, 
\qquad 
T=g^{\mu\nu}T_{\mu\nu}.
\end{equation*}

We adopt the effective anisotropic-fluid Lagrangian
\begin{equation*}
L_m^{\mathrm{eff}}=p_t-\frac{1}{2}(\rho+p_t)(g^{\alpha\beta}u_\alpha u_\beta+1)
-\frac{1}{2}(p_r-p_t)(g^{\alpha\beta}\chi_\alpha\chi_\beta-1),
\end{equation*}
with
\begin{equation*}
u_\mu u^\mu=-1, \qquad \chi_\mu\chi^\mu=1, \qquad u_\mu\chi^\mu=0.
\end{equation*}

For Lagrangians without metric derivatives,
\begin{equation*}
T_{\mu\nu}=g_{\mu\nu}L_m-2\frac{\partial L_m}{\partial g^{\mu\nu}},
\end{equation*}
which yields
\begin{equation*}
T_{\mu\nu}=(\rho+p_t)u_\mu u_\nu+p_t g_{\mu\nu}+(p_r-p_t)\chi_\mu\chi_\nu.
\end{equation*}

The tensor $\Theta_{\mu\nu}$ is
\begin{equation*}
\Theta_{\mu\nu}=g^{\alpha\beta}\frac{\delta T_{\alpha\beta}}{\delta g^{\mu\nu}}
=-2T_{\mu\nu}+p_t g_{\mu\nu}.
\end{equation*}

For the quadratic model
\begin{equation*}
f(R,T)=R+\alpha T^2,
\end{equation*}
we have
\begin{equation*}
f_R=1, \qquad f_T=2\alpha T,
\end{equation*}
so all derivative terms of $f_R$ vanish. The field equations reduce to
\begin{equation}
R_{\mu\nu}-\frac{1}{2}(R+\alpha T^2)g_{\mu\nu}
=8\pi T_{\mu\nu}+2\alpha T(T_{\mu\nu}+\Theta_{\mu\nu}).
\end{equation}

\subsection{Background geometry and perturbation ansatz}

The static spherically symmetric metric is
\begin{equation*}
ds^2=-e^{2\mu_0(r)}dt^2+e^{2\nu_0(r)}dr^2+r^2(d\theta^2+\sin^2\theta\,d\phi^2).
\end{equation*}

The background stress tensor is
\begin{equation*}
T^{(0)}_{\mu\nu}=(\rho_0+p_{t0})u_\mu u_\nu+p_{t0}g_{\mu\nu}+(p_{r0}-p_{t0})\chi_\mu\chi_\nu,
\end{equation*}
with trace
\begin{equation*}
T_0=-\rho_0+p_{r0}+2p_{t0}.
\end{equation*}

We perturb the metric as
\begin{equation*}
g_{\mu\nu}=g^{(0)}_{\mu\nu}+h_{\mu\nu}.
\end{equation*}

In Regge--Wheeler gauge, axial perturbations are
\begin{equation*}
h_{t\phi}=h_0(r)e^{-i\omega t}S_\ell(\theta), \qquad
h_{r\phi}=h_1(r)e^{-i\omega t}S_\ell(\theta),
\end{equation*}
with
\begin{equation*}
S_\ell(\theta)=\sin\theta\,\partial_\theta P_\ell(\cos\theta).
\end{equation*}

The axial fluid perturbation is
\begin{equation*}
\delta u_\phi=r^2\varpi(r)e^{-i\omega t}S_\ell.
\end{equation*}

The only nonzero perturbed stress components are
\begin{align*}
\delta T_{t\phi}&=\Bigl[-(\rho_0+p_{t0})r^2\varpi+p_{t0}h_0\Bigr]e^{-i\omega t}S_\ell,\\
\delta T_{r\phi}&=p_{t0}h_1e^{-i\omega t}S_\ell.
\end{align*}

\subsection{Vanishing of $\delta T$, $\delta f_T$, and $\delta L_m^{\mathrm{eff}}$}

The trace perturbation is
\begin{equation*}
\delta T=\delta(g^{\mu\nu}T_{\mu\nu})
=-h^{\mu\nu}T^{(0)}_{\mu\nu}+g^{(0)\mu\nu}\delta T_{\mu\nu}.
\end{equation*}

Both terms vanish in the odd sector, hence
\begin{equation*}
\delta T=0,
\qquad
\delta f_T=0.
\end{equation*}

Similarly,
\begin{equation*}
\delta L_m^{\mathrm{eff}}=0.
\end{equation*}

These identities are the key simplifications of the axial sector.

\subsection{Explicit variation of the $f_T$-sector}

We compute
\begin{equation*}
\delta\bigl[f_T(T_{\mu\nu}+\Theta_{\mu\nu})\bigr].
\end{equation*}

Since $\delta f_T=0$,
\begin{equation*}
\delta[\cdots]=2\alpha T_0(\delta T_{\mu\nu}+\delta\Theta_{\mu\nu}).
\end{equation*}

Using
\begin{equation*}
\delta\Theta_{\mu\nu}=-2\delta T_{\mu\nu}+p_{t0}h_{\mu\nu},
\end{equation*}
we obtain
\begin{align*}
\delta T_{t\phi}+\delta\Theta_{t\phi}&=(\rho_0+p_{t0})r^2\varpi e^{-i\omega t}S_\ell,\\
\delta T_{r\phi}+\delta\Theta_{r\phi}&=0.
\end{align*}

The trace-coupling therefore contributes only to the $t\phi$ equation.

\subsection{Linearized odd-parity field equations}

Linearizing the field equations yields
\begin{equation*}
\delta R_{\mu\nu}-\frac{1}{2}(R_0+\alpha T_0^2)h_{\mu\nu}
=8\pi \delta T_{\mu\nu}+2\alpha T_0(\delta T_{\mu\nu}+\delta\Theta_{\mu\nu}).
\end{equation*}

Using the Ricci perturbations in Regge--Wheeler gauge then yields two coupled equations for $h_0$ and $h_1$ before $\varpi$ is eliminated.

\subsection{Linearized modified divergence law}

The matter equation follows from
\begin{equation*}
\nabla_\mu T^\mu_{\ \nu}
=\frac{\alpha}{4\pi-\alpha T}\left[T^\mu_{\ \nu}\partial_\mu T
-T\partial_\nu p_t-\left(p_t+\frac{1}{2}T\right)\partial_\nu T\right].
\end{equation*}

Linearizing the $\nu=\phi$ component yields
\begin{equation*}
-i\omega(\rho_0+p_{t0})\varpi+\Xi_0(r)h_1=0,
\end{equation*}
with
\begin{equation*}
\Xi_0(r)=\frac{p_{t0}-p_{r0}}{r}+(\rho_0+p_{t0})'
-\frac{2\alpha p_{t0}T_0'}{8\pi-2\alpha T_0}.
\end{equation*}

Hence
\begin{equation*}
\varpi=\frac{\Xi_0(r)}{i\omega(\rho_0+p_{t0})}h_1.
\end{equation*}

\subsection{Closed two-field system}

Substituting $\varpi$ into the gravitational equations yields
\begin{align*}
h_0''+A_1 h_0'+A_2 h_0+A_3 h_1'+A_4 h_1&=0,\\
h_1''+B_1 h_1'+B_2 h_1+B_3 h_0'+B_4 h_0&=0,
\end{align*}

with coefficients given explicitly in Sec.~II.

These are the equations solved by the PINN. On static admissible branches the same odd sector also admits the exact gauge-invariant master-equation reduction summarized in Appendix~\ref{app:master_exact}, which is the production variable used for the negative-$w_r$ scan.

\subsection{Closure of the odd sector}

The odd sector contains three amplitudes
\begin{equation*}
h_0(r), \qquad h_1(r), \qquad \varpi(r).
\end{equation*}

The two gravitational equations and the modified divergence law form a closed system. Eliminating $\varpi$ yields the two-field system quoted above.

\subsection{Consistency checks}

Two limits provide important checks.

\medskip
\noindent\textbf{GR limit ($\alpha\to0$):}
\begin{equation*}
\nabla_\mu T^\mu_{\ \nu}=0,
\end{equation*}
and the system reduces to the GR axial equations.

\medskip
\noindent\textbf{Vacuum limit:}
\begin{equation*}
\rho_0=p_{r0}=p_{t0}=T_0=0,
\end{equation*}
so all matter-dependent terms vanish and the system reduces to the Regge--Wheeler equation.

\medskip
In summary, $f_R=1$ removes all derivative terms exactly, $\delta T=0$ does not eliminate the $f_T$ sector because it still contributes through $2\alpha T_0(\delta T_{\mu\nu}+\delta\Theta_{\mu\nu})$, and the axial matter equation follows from the modified divergence law rather than from GR conservation.

\section{Background solver, horizon regularity, and asymptotic admissibility}
\label{app:background}

This appendix collects the derivations underlying the background workflow summarized in the main text.

\subsection{Near-horizon index and regularity window}

Let $r=r_H+\Delta r$ with $\Delta r\ll r_H$ and assume a nontrivial matter-supported branch,
\begin{equation}
N(r)=N_1\Delta r+\cdots,
\qquad
\sigma(r)=\sigma_H+\cdots,
\qquad
\rho_0(r)=\rho_s\,\Delta r^s+\cdots.
\end{equation}
Regularity of the horizon implies $N_1=1/r_H$ once $\rho_0\to0$ there. Using Eq.~\eqref{eq:rhoprime} and keeping the leading $1/\Delta r$ term from $N'/(2N)$ gives
\begin{equation}
\frac{\rho_0'}{\rho_0}= -\frac{1+w_r}{2w_r}\frac{1}{\Delta r}+O(1).
\end{equation}
Integrating immediately yields the indicial exponent
\begin{equation}
s=-\frac{1+w_r}{2w_r}.
\end{equation}
Hence nontrivial decaying matter support requires $w_r<0$. The additional requirement that $\sigma'/\sigma$ remain finite at the horizon implies $s\ge1$, which narrows the regular branch to
\begin{equation}
-\frac{1}{3}\le w_r<0.
\end{equation}
This is the origin of Eq.~\eqref{eq:wr_constraint} in the main text.

\subsection{Asymptotic index and finite-mass condition}

Assume that the matter profile decays as
\begin{equation}
\rho_0(r)\sim \rho_\infty r^{-n_\infty},
\qquad r\to\infty,
\end{equation}
with $N\to1$ and $\sigma\to1$. In Eq.~\eqref{eq:rhoprime}, the terms involving $N'/N$ and $\sigma'/\sigma$ are then subleading compared with the explicit anisotropy term, and one obtains
\begin{equation}
\frac{\rho_0'}{\rho_0}=-\frac{2(w_r-w_t)}{w_r}\frac{1}{r}+O\!\left(r^{1-n_\infty}\right),
\end{equation}
so that
\begin{equation}
n_\infty=\frac{2(w_r-w_t)}{w_r}.
\end{equation}
Because the effective mass function satisfies $m'(r)\sim 4\pi r^2\rho_0$, finite ADM mass requires $n_\infty>3$. For negative $w_r$, this gives the tangential-pressure condition
\begin{equation}
w_t>-\frac{w_r}{2}.
\end{equation}
The trace then falls as $T_0=q\rho_0\sim r^{-n_\infty}$, not as $r^{-3}$ in general.

\subsection{Illustrative integration on an admissible background}

\begin{figure}[t]
\centering
\includegraphics[width=0.98\columnwidth]{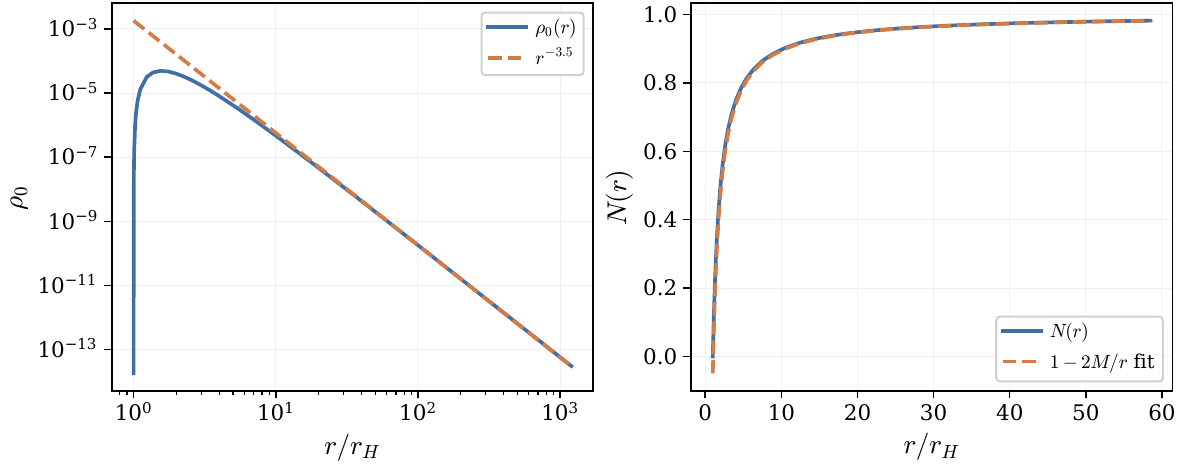}
\caption{Independent outward integration on an admissible background, $(w_r,w_t,\alpha)=(-0.2,0.15,0)$ with $r_H=1$ and $\rho_s=0.01835620783535$. Left: the matter profile $\rho_0(r)$ together with the reference power law $r^{-3.5}$ predicted by Eq.~\eqref{eq:ninfty}. Right: the metric function $N(r)$ and its asymptotic approach to $1-2M/r$. The fitted exponent $n_{\rm fit}=3.497$ agrees with the analytic prediction $n_\infty=3.5$.}
\label{fig:background_asymptotics}
\end{figure}

To show explicitly that the admissible sector is nonempty, we performed an independent outward integration of Eqs.~\eqref{eq:Nprime}--\eqref{eq:rhoprime} for the representative case
\begin{equation}
(w_r,w_t,\alpha)=(-0.2,0.15,0),
\qquad r_H=1,
\qquad \rho_s=0.01835620783535.
\end{equation}
For this choice the analytic predictions are $s=2$ and $n_\infty=3.5$. We integrated from $r=1+10^{-6}$ to $r=2500$ with a high-accuracy adaptive DOP853 solver, holding the anchored near-horizon coefficient fixed and fitting the asymptotic tail on the far exterior. The resulting metrics are
\begin{equation}
s=2,
\qquad
n_{\rm fit}=3.497,
\qquad
M/r_H\approx0.624628.
\end{equation}
Figure~\ref{fig:background_asymptotics} shows the corresponding density decay and the approach of $N(r)$ to its asymptotic Schwarzschild form.

\section{Weak-coupling frequency shift and its discrete implementation}
\label{app:weakcoupling}

The main text uses the first-order correction $\Omega_1$ in the small-$\alpha$ comparison. For completeness, we derive it here once the compactified operator is organized as
\begin{equation}
\mathcal{L}(\Omega,\alpha)\mathbf{g}=0,
\qquad
\mathbf{g}(x)=\begin{pmatrix}g_0(x)\\ g_1(x)\end{pmatrix},
\end{equation}
with the exact horizon regularity already built into the ansatz for $\mathbf{g}$. Expand the operator, the eigenfunction, and the eigenvalue around the Schwarzschild point,
\begin{align}
\mathcal{L}(\Omega,\alpha)&=\mathcal{L}_0(\Omega)+\alpha\,\mathcal{L}_1(\Omega)+O(\alpha^2),
\\
\mathbf{g}&=\mathbf{g}_0+\alpha\,\mathbf{g}_1+O(\alpha^2),
\\
\Omega&=\Omega_0+\alpha\,\Omega_1+O(\alpha^2),
\end{align}
where $\Omega_0=\Omega_{\Schw}$ and $\mathcal{L}_0(\Omega_0)\mathbf{g}_0=0$. The $O(\alpha)$ equation is
\begin{equation}
\mathcal{L}_0(\Omega_0)\mathbf{g}_1+\left[(\partial_\Omega\mathcal{L}_0)_{\Omega_0}\,\Omega_1+\mathcal{L}_1(\Omega_0)\right]\mathbf{g}_0=0.
\label{eq:firstorder_operator_eq}
\end{equation}
Because the QNM problem is non-self-adjoint, the solvability condition is imposed with the adjoint zero mode $\boldsymbol{\chi}_0$ satisfying
\begin{equation}
\mathcal{L}_0^{\dagger}(\Omega_0)\boldsymbol{\chi}_0=0.
\end{equation}
Using the bilinear pairing
\begin{equation}
\langle \boldsymbol{\chi},\mathbf{u}\rangle\equiv\int_0^1 \boldsymbol{\chi}^{\dagger}(x)\cdot\mathbf{u}(x)\,w(x)\,\dd x,
\end{equation}
with any convenient positive weight $w(x)$ on the compactified interval, Eq.~\eqref{eq:firstorder_operator_eq} yields the Fredholm condition
\begin{equation}
\Omega_1=-\frac{\left\langle \boldsymbol{\chi}_0,\mathcal{L}_1(\Omega_0)\mathbf{g}_0\right\rangle}{\left\langle \boldsymbol{\chi}_0,(\partial_\Omega\mathcal{L}_0)_{\Omega_0}\mathbf{g}_0\right\rangle}.
\label{eq:omega1_formal_appendix}
\end{equation}
Equation~\eqref{eq:omega1_formal_appendix} is the formal weak-coupling correction quoted in the main text.

In practice the external validator works with a Chebyshev collocation discretization. If the discretized problem is written as
\begin{equation}
\mathbf{A}(\Omega,\alpha)\mathbf{v}=0,
\end{equation}
then the corresponding first-order correction is
\begin{equation}
\Omega_1=-\frac{\mathbf{u}_0^{\dagger}\,\mathbf{A}_1(\Omega_0)\,\mathbf{v}_0}{\mathbf{u}_0^{\dagger}(\partial_\Omega \mathbf{A}_0)_{\Omega_0}\mathbf{v}_0},
\label{eq:omega1_matrix}
\end{equation}
where $\mathbf{v}_0$ and $\mathbf{u}_0$ are the right and left null vectors of $\mathbf{A}_0(\Omega_0)$. The weak-coupling comparison used in the main text can therefore be reproduced either from the continuum operator via Eq.~\eqref{eq:omega1_formal_appendix} or directly from the matrix discretization via Eq.~\eqref{eq:omega1_matrix}.

Two points deserve emphasis. First, the correction includes both the explicit perturbative change of the compactified operator and the first-order change in the background profiles. Second, because the odd sector is genuinely coupled, $\mathbf{g}_0$ and $\boldsymbol{\chi}_0$ are two-component objects even when the Schwarzschild limit reduces to the vacuum Regge--Wheeler problem in a particular basis.

\section{Explicit time-domain odd system, principal symbol, and Chebyshev matrix formulation}
\label{app:principal}

\subsection{Unreduced time-domain odd system}

Before Fourier transformation and before eliminating the axial matter amplitude, the odd-parity variables $h_0(t,r)$, $h_1(t,r)$, and $\varpi(t,r)$ satisfy
\begin{align}
0&=\mathcal{E}_{t\phi}\equiv h_{0,rr}+A_1 h_{0,r}-e^{2(\nu_0-\mu_0)}h_{0,tt}-h_{1,tr}-A_1 h_{1,t}+V_0(r)h_0-C_{\varpi}(r)\varpi,
\label{eq:td_h0_appendix}
\\
0&=\mathcal{E}_{r\phi}\equiv h_{1,rr}+B_1 h_{1,r}-e^{2(\nu_0-\mu_0)}h_{1,tt}-h_{0,tr}+\nu_0' h_{0,t}+V_1(r)h_1,
\label{eq:td_h1_appendix}
\\
0&=\mathcal{E}_{\phi}\equiv (\rho_0+p_{t0})\,\varpi_{,t}+\Xi_0(r)h_1,
\label{eq:td_varpi_appendix}
\end{align}
with
\begin{align}
A_1&=\frac{2}{r}-\nu_0'+\mu_0',
&
B_1&=\frac{2}{r}+\mu_0'-\nu_0',
\\
V_0(r)&=e^{2\nu_0}\bigl(R_0+\alpha T_0^2+16\pi p_{t0}\bigr),
&
V_1(r)&=e^{2\nu_0}\left[\frac{2-\ell(\ell+1)}{r^2}+R_0+\alpha T_0^2+16\pi p_{t0}\right],
\\
C_{\varpi}(r)&=2e^{2\nu_0}r^2\bigl(8\pi-2\alpha T_0\bigr)(\rho_0+p_{t0}),
&
\Xi_0(r)&=\frac{p_{t0}-p_{r0}}{r}+(\rho_0+p_{t0})'-\frac{2\alpha p_{t0}T_0'}{8\pi-2\alpha T_0}.
\end{align}
The frequency-domain equations in Section~\ref{subsec:oddsector} follow after inserting $h_a(t,r)=h_a(r)e^{-i\omega t}$ and $\varpi(t,r)=\varpi(r)e^{-i\omega t}$ into Eqs.~\eqref{eq:td_h0_appendix}--\eqref{eq:td_varpi_appendix} and then eliminating $\varpi$ with Eq.~\eqref{eq:td_varpi_appendix}. This makes explicit that the axial matter amplitude enters locally in time and that the $1/(i\omega)$ term appears only after frequency-domain elimination.

\subsection{Principal-symbol extraction}

The hyperbolicity scan is performed on the unreduced time-domain system above rather than on the reduced frequency-domain pair. In practice, the equations are organized schematically as
\begin{equation}
E_A\equiv K_{AB}(r)\,\partial_t^2 q_B+2M_{AB}(r)\,\partial_t\partial_r q_B-P_{AB}(r)\,\partial_r^2 q_B+Q_{AB}(r)\,\partial_t q_B+R_{AB}(r)\,\partial_r q_B+S_{AB}(r)q_B=0,
\label{eq:generic_principal}
\end{equation}
with $q_A=(h_0,h_1,\varpi)$ after the standard first-order companion reduction of the matter equation. The pure second-order gravitational block is read off directly from Eqs.~\eqref{eq:td_h0_appendix} and \eqref{eq:td_h1_appendix}; in the sign convention of Eq.~\eqref{eq:generic_principal} one finds
\begin{equation}
\mathbf{K}_{(hh)}=-e^{2(\nu_0-\mu_0)}\begin{pmatrix}1&0\\0&1\end{pmatrix},
\qquad
\mathbf{M}_{(hh)}=-\frac{1}{2}\begin{pmatrix}0&1\\1&0\end{pmatrix},
\qquad
\mathbf{P}_{(hh)}=-\begin{pmatrix}1&0\\0&1\end{pmatrix},
\label{eq:principal_block_appendix}
\end{equation}
up to an overall sign that does not affect the characteristic equation. The remaining rows and columns are supplied by the companion reduction of Eq.~\eqref{eq:td_varpi_appendix} and by the chosen state vector used in the numerical scan. For a local WKB mode,
\begin{equation}
q_B\propto \hat q_B\exp\bigl[-i\omega t+ik(r-r_0)\bigr],
\end{equation}
Eq.~\eqref{eq:generic_principal} becomes
\begin{equation}
\left[-\omega^2\mathbf{K}-2\omega k\mathbf{M}+k^2\mathbf{P}+O(k)\right]\hat{\mathbf{q}}=0.
\end{equation}
Defining the phase speed $c=\omega/k$, the characteristic equation is
\begin{equation}
\det\bigl[c^2\mathbf{K}+2c\mathbf{M}-\mathbf{P}\bigr]=0.
\label{eq:characteristic_det_appendix}
\end{equation}
The hyperbolicity scan used in this work is:
\begin{enumerate}
\item assemble $\mathbf{K}(r_i)$, $\mathbf{M}(r_i)$, and $\mathbf{P}(r_i)$ on the same radial grid used for the background interpolation,
\item solve Eq.~\eqref{eq:characteristic_det_appendix} for all characteristic speeds $c_A(r_i)$,
\item discard roots with non-negligible imaginary part and mark the point as non-hyperbolic if any required branch becomes complex,
\item sort the remaining real branches and define $c_-^2(r_i)$ as the smallest physical branch,
\item compute $\minH=\min_i c_-^2(r_i)$ and apply the practical acceptance cut $\minH>10^{-4}$.
\end{enumerate}
This procedure is the one used to construct the admissibility diagnostic in the main text.

\subsection{Chebyshev polynomial eigenvalue problem}

Let $x_j=\cos(j\pi/N_{\Cheb})$ be the Chebyshev--Gauss--Lobatto nodes mapped to the compactified radial interval. Discretizing Eq.~\eqref{eq:compactified_operator} with the usual differentiation matrices produces a quadratic polynomial eigenvalue problem in $z=i\Omega$,
\begin{equation}
\bigl[\mathbf{A}_0+z\mathbf{A}_1+z^2\mathbf{A}_2\bigr]\mathbf{v}=0.
\end{equation}
A standard linearization introduces the doubled state $\mathbf{y}=(\mathbf{v},z\mathbf{v})^T$ and yields
\begin{equation}
\begin{pmatrix}
\mathbf{0} & \mathbf{I} \\
-\mathbf{A}_0 & -\mathbf{A}_1
\end{pmatrix}
\mathbf{y}
=z
\begin{pmatrix}
\mathbf{I} & \mathbf{0} \\
\mathbf{0} & \mathbf{A}_2
\end{pmatrix}
\mathbf{y}.
\label{eq:linearized_qep}
\end{equation}
The physical QNM candidates are those eigenvalues $z$ whose reconstructed eigenvectors satisfy the same horizon normalization and asymptotic regularity as the PINN solution. Equation~\eqref{eq:linearized_qep} is the deterministic matrix problem used for the independent low-coupling cross-check and provides the natural route for stronger-coupling validation.

\subsection{Strong-shift validation protocol}

For the strong-coupling regime, we adopt the following validation criteria:
\begin{enumerate}
\item the background satisfies the admissibility conditions of Appendix~B,
\item the principal-symbol diagnostic gives $\minH>10^{-4}$ on the full radial interval,
\item the Chebyshev eigenvalue is stable when $N_{\Cheb}$ is increased, for example from $N_{\Cheb}=100$ to $150$ and $200$,
\item the PINN and Chebyshev frequencies agree within the combined conservative uncertainty envelope whenever both calculations are available,
\item the mode track is continuous under parameter continuation from a neighboring validated point.
\end{enumerate}
This protocol defines the minimum standard for any future full coupled-PINN strong-shift calculation in the admissible negative-$w_r$ sector.

\section{Exact odd-sector reduction on the admissible branch}
\label{app:master_exact}

Starting from Eq.~\eqref{eq:einsteinlike}, define
\begin{equation}
T_{\mu\nu}^{\rm eff}=c\,T_{\mu\nu}+d\,g_{\mu\nu},
\qquad
c=1-\frac{\alpha T}{4\pi},
\qquad
d=\frac{\alpha}{8\pi}\left(2p_tT+\frac{1}{2}T^2\right).
\end{equation}
Then the quadratic trace model can be written exactly as
\begin{equation}
G_{\mu\nu}=8\pi T_{\mu\nu}^{\rm eff}.
\end{equation}
For the anisotropic-fluid form used in the main text this implies
\begin{equation}
\rho_{\rm eff}=c\rho-d,
\qquad
p_{r,\rm eff}=cp_r+d,
\qquad
p_{t,\rm eff}=cp_t+d.
\end{equation}
In odd parity the scalar fluid variables do not fluctuate, so
\begin{equation}
\delta\rho=\delta p_r=\delta p_t=\delta T=\delta c=\delta d=0.
\end{equation}
The odd-parity problem is therefore exactly the odd-parity problem of Einstein gravity on a static background sourced by the frozen effective anisotropic fluid.

For a static line element
\begin{equation}
\dd s^2=-A(r)\dd t^2+\frac{\dd r^2}{B(r)}+C(r)\dd\Omega_2^2,
\end{equation}
the gauge-invariant master variable $\Psi$ obeys
\begin{equation}
\frac{\dd^2\Psi}{\dd r_*^2}+\bigl[\omega^2-V_{\rm ax}(r)\bigr]\Psi=0,
\qquad
\frac{\dd r_*}{\dd r}=\frac{1}{\sqrt{AB}},
\end{equation}
with
\begin{equation}
V_{\rm ax}(r)=\frac{[\ell(\ell+1)-2]A}{C}
+\sqrt{ABC}\,\frac{\dd}{\dd r}\left[\sqrt{AB}\,\frac{\dd}{\dd r}\left(C^{-1/2}\right)\right].
\end{equation}
For the metric of Eq.~\eqref{eq:Nsigma_metric},
\begin{equation}
\Psi=N\sigma h_1,
\qquad
V_{\rm ax}(r)=\frac{[\ell(\ell+1)-2]N\sigma^2}{r^2}
-\frac{N\sigma}{r}\frac{\dd(N\sigma)}{\dd r}
+\frac{2(N\sigma)^2}{r^2}.
\end{equation}
In vacuum one has $\sigma=1$ and $N=1-r_H/r$, and the potential reduces exactly to the Regge--Wheeler form. This is the exact production equation used for the admissible-branch scan reported in the main text.

\section{Positive-$w_r$ comparison catalogues and auxiliary anisotropy scan}
\label{app:comparisoncatalogues}

To keep the main text focused on the admissible branch, we collect here the pointwise positive-$w_r$ comparison catalogue and the auxiliary anisotropy scan used only for numerical benchmarking. Table~\ref{tab:alpha_scan} lists the comparison-branch frequencies and validation statuses used in Sec.~4 only as numerical benchmarks.

\begin{table*}[t]
\centering
\small
\setlength{\tabcolsep}{5pt}
\renewcommand{\arraystretch}{1.08}
\caption{Frequency catalogue for the positive-$w_r$ comparison branch with $w_r=0.2$ and $w_t=0.15$. Numbers in parentheses denote seed-to-seed scatter only; the broader conservative uncertainty budget is summarized in Table~\ref{tab:uncertainty}. The status column identifies which points are externally validated, internally validated, provisional, or marginal. Because this branch violates the regularity constraints of Section~\ref{subsec:background}, the entries beyond the Schwarzschild point are included only for numerical comparison.}
\label{tab:alpha_scan}
\begin{tabular*}{\textwidth}{@{\extracolsep{\fill}}cccccc@{}}
\toprule
$\alpha/M^2$ & $\Omega=r_H\omega$ & $\Ohat=M\omega$ & $\minH$ & $L_{\tot}$ & status \\
\midrule
0.000 & $0.74730(5)-0.17792(3)i$ & $0.37365-0.08896i$ & 0.4620 & $1.6\times10^{-6}$ & \shortstack[l]{externally\\validated} \\
0.001 & $0.74724(7)-0.17784(5)i$ & $0.37362-0.08892i$ & 0.4580 & $5.0\times10^{-5}$ & \shortstack[l]{externally\\validated} \\
0.010 & $0.74666(12)-0.17687(8)i$ & $0.37333-0.08843i$ & 0.4410 & $2.8\times10^{-3}$ & \shortstack[l]{externally\\validated} \\
0.050 & $0.74308(24)-0.17234(15)i$ & $0.37165-0.08620i$ & 0.3570 & $6.6\times10^{-3}$ & \shortstack[l]{externally\\validated} \\
0.100 & $0.73678(40)-0.16596(27)i$ & $0.36861-0.08303i$ & 0.2740 & $1.02\times10^{-2}$ & \shortstack[l]{externally\\validated} \\
0.150 & $0.72844(64)-0.15851(44)i$ & $0.36455-0.07933i$ & 0.1920 & $1.8\times10^{-2}$ & \shortstack[l]{internally\\validated} \\
0.200 & $0.71818(92)-0.14998(66)i$ & $0.35956-0.07509i$ & 0.1160 & $3.1\times10^{-2}$ & \shortstack[l]{internally\\validated} \\
0.250 & $0.70596(130)-0.14036(90)i$ & $0.35358-0.07030i$ & 0.0550 & $5.4\times10^{-2}$ & \shortstack[l]{internally\\validated} \\
0.300 & $0.69205(186)-0.12956(122)i$ & $0.34679-0.06492i$ & 0.0180 & $8.7\times10^{-2}$ & provisional \\
0.325 & $0.6841(24)-0.1237(16)i$ & $0.34291-0.06200i$ & 0.0048 & $1.15\times10^{-1}$ & marginal \\
0.350 & --- & --- & $-0.0060$ & $>1$ & inadmissible \\
\bottomrule
\end{tabular*}
\end{table*}

\subsection{Auxiliary anisotropy scan at fixed \texorpdfstring{$\alpha/M^2=0.1$}{alpha/M^2=0.1}}

To isolate the role of anisotropy in the comparison scan, we fix $\alpha/M^2=0.1$ and vary $(w_r,w_t)$. The quantity
\begin{equation}
\Ag\equiv\frac{\|g_0\|_2}{\|g_1\|_2},
\qquad
\|g_a\|_2^2\equiv\int_0^1|g_a(x)|^2\,\dd x,
\label{eq:Ag_def_main}
\end{equation}
is described only as a basis-dependent admixture indicator. Because the homogeneous problem is normalized by fixing $g_0(1)=1$ and $g_1(1)=\Lambda_H(\Omega)$, $\Ag$ is not a physical energy fraction.

\begin{table*}[t]
\centering
\small
\setlength{\tabcolsep}{6pt}
\renewcommand{\arraystretch}{1.08}
\caption{Dependence of the fundamental axial $\ell=2$ mode on pressure anisotropy at fixed $\alpha/M^2=0.1$ along the positive-$w_r$ comparison branch. The last column is the basis-dependent regular-field admixture diagnostic $\Ag$ defined in Eq.~\eqref{eq:Ag_def_main}. These entries are included for numerical comparison rather than as astrophysical predictions.}
\label{tab:anisotropy}
\begin{tabular*}{0.78\textwidth}{@{\extracolsep{\fill}}cccc@{}}
\toprule
$w_r$ & $w_t$ & $\Omega$ & $\Ag$ \\
\midrule
0.1 & 0.1  & $0.74017(35)-0.16849(24)i$ & 0.034 \\
0.2 & 0.1  & $0.73908(40)-0.16731(28)i$ & 0.049 \\
0.3 & 0.1  & $0.73790(45)-0.16606(32)i$ & 0.066 \\
0.1 & 0.15 & $0.73886(38)-0.16705(26)i$ & 0.060 \\
0.2 & 0.15 & $0.73678(40)-0.16596(27)i$ & 0.089 \\
0.3 & 0.15 & $0.73405(47)-0.16458(34)i$ & 0.115 \\
0.1 & 0.2  & $0.73606(44)-0.16524(30)i$ & 0.086 \\
0.2 & 0.2  & $0.73298(52)-0.16342(36)i$ & 0.126 \\
0.3 & 0.2  & $0.72973(60)-0.16150(42)i$ & 0.161 \\
0.1 & 0.3  & $0.72908(58)-0.16106(40)i$ & 0.146 \\
0.2 & 0.3  & $0.72475(72)-0.15833(48)i$ & 0.201 \\
0.3 & 0.3  & $0.72027(88)-0.15549(56)i$ & 0.251 \\
\bottomrule
\end{tabular*}
\end{table*}

\begin{figure}[t]
\centering
\includegraphics[width=0.98\columnwidth]{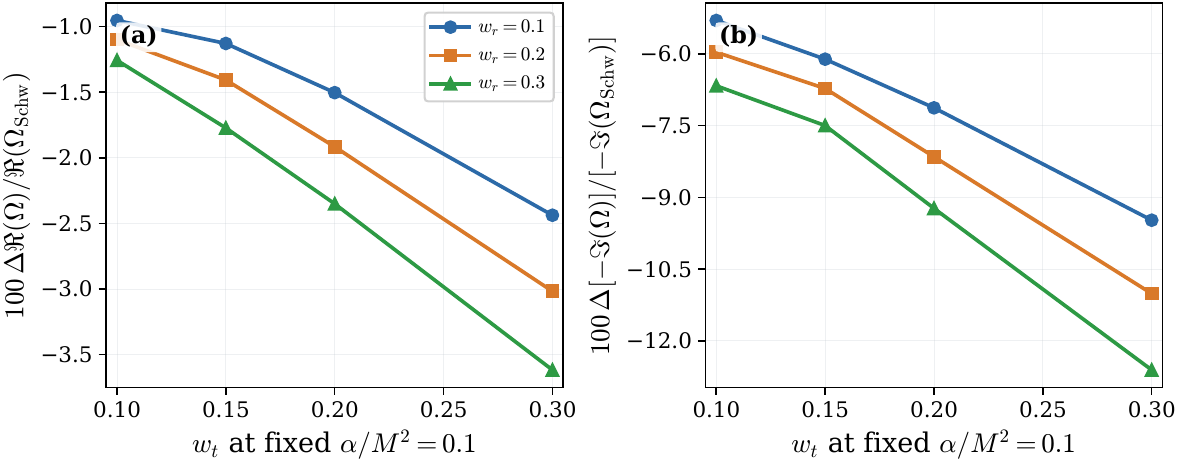}
\caption{Percentage frequency shifts in the anisotropy scan on the positive-$w_r$ comparison branch at fixed $\alpha/M^2=0.1$ as a function of $w_t$ for three values of $w_r$. The left panel shows $100\,\Delta\Re(\Omega)/\Re(\Omega_\mathrm{Schw})$, and the right panel shows $100\,\Delta[-\Im(\Omega)]/[-\Im(\Omega_\mathrm{Schw})]$. The steeper dependence on $w_t$ quantifies the stronger sensitivity to tangential pressure; the branch is shown here only for numerical comparison.}
\label{fig:anisotropy}
\end{figure}

Figure~\ref{fig:anisotropy} makes the trend clear. Increasing $w_t$ produces a much larger spectral shift than changing $w_r$ by a comparable amount. For example, at fixed $w_r=0.1$, moving from $w_t=0.1$ to $w_t=0.3$ lowers $\Re(\Omega)$ by about $2.44\%$ and $-\Im(\Omega)$ by about $9.48\%$ relative to Schwarzschild. At fixed $w_t=0.1$, changing $w_r$ from $0.1$ to $0.3$ alters $\Re(\Omega)$ by only about $0.30\%$ and $-\Im(\Omega)$ by about $1.37\%$. This asymmetry is consistent with the coefficient structure in Eq.~\eqref{eq:ABcoeffs_main}, where the combinations $\rho_0+p_{t0}$, $p_{t0}$, and $p_{t0}'$ enter directly into the coupling terms.

\end{document}